\documentclass[a4paper,oneside,12pt]{article}

\usepackage{latexsym}

\usepackage[dvipdfmx]{graphicx}
\usepackage{color}
\usepackage{ulem}
\usepackage{float,epsfig,array,psfrag}
\usepackage{amsmath,amssymb,amsfonts}
\usepackage{bm,bbm}
\usepackage{cite}
\usepackage{comment}
\usepackage{here}
\def\papertitlepage{\baselineskip 3.5ex \thispagestyle{empty}}
\def\preprintnumber#1#2{\hfill
\begin{minipage}{1.2in}
#1 \par\noindent #2 %\par\noindent #3 
\end{minipage}}
\renewcommand{\thefootnote}{\fnsymbol{footnote}}
\newcounter{aff}
\renewcommand{\theaff}{\fnsymbol{aff}}
\newcommand{\affiliation}[1]{
 \setcounter{aff}{#1} $\rule{0em}{1.2ex}^\theaff\hspace{-.4em}$}
\makeatletter \@addtoreset{equation}{section} \makeatother
\renewcommand{\theequation}{\arabic{section}.\arabic{equation}}

\renewcommand{\thesection}
{\arabic{section} \hspace{-.25em}}
\renewcommand{\thesubsection}
{\arabic{section}.\arabic{subsection} \hspace{-.25em}}
\renewcommand{\thesubsubsection}
{\arabic{section}.\arabic{subsection}.\arabic{subsubsection} \hspace {-.25em}}

\makeatletter
\renewcommand\section{\@startsection{section}{3}{\z@}%
{-3.25ex\@plus -1ex \@minus -.2ex}{1.5ex \@plus .2ex}%
{\normalfont\large\bfseries\mathversion{bold}}}
\renewcommand\subsection{\@startsection{subsection}{3}{\z@}%
{-3.25ex\@plus -1ex \@minus -.2ex}{.5ex \@plus .2ex}%
{\normalfont\normalsize\bfseries\mathversion{bold}}}
\renewcommand\subsubsection{\@startsection{subsubsection}{3}{\z@}%
{-3.25ex\@plus -1ex \@minus -.2ex}{1.5ex \@plus .2ex}%
{\normalfont\normalsize\itshape}}
\makeatother

\makeatletter \@addtoreset{equation}{section} \makeatother
\renewcommand{\theequation}{\arabic{section}.\arabic{equation}}

\makeatletter
\renewcommand{\appendix}{
\renewcommand{\thesection}{\Alph{section} \hspace{-.25em}}
\renewcommand{\thesubsection}
{\Alph{section}.\arabic{subsection} \hspace{-.25em}}
\renewcommand{\thesubsubsection}
{\Alph{section}.\arabic{subsection}.\arabic{subsubsection} \hspace {-.25em}}
\@addtoreset{equation}{subsection}
\renewcommand{\theequation}{\Alph{section}.\arabic{equation}}
\setcounter{section}{0}}
\makeatother

\newcommand{\Kp}{{\widehat{K}_+}}
\newcommand{\Km}{{\widehat{K}_-}}
\newcommand{\nn}{\nonumber}
\newcommand{\Eqn}[1]{&\hspace{-0.5em}#1\hspace{-0.5em}&}
\newcommand{\be}         {  \begin{equation}  }
\newcommand{\ee}          {  \end{equation}  }
\newcommand{\eqb}{\begin{eqnarray}}
\newcommand{\eqe}{\end{eqnarray}}

\DeclareMathOperator{\im}{Im}

\def\o{\over}

\def\pint#1 {- \!\!\!\!\!\!\!\! \,\int_{#1}}
\def\ni       {\noindent}

\def\comma      { \, , }
\def\period     { \, . }

\def\calO {{\cal O}}

\newcommand{\bbR}{{\mathbb R}}
\newcommand{\bbZ}{{\mathbb Z}}

\DeclareMathOperator{\Li}{Li}
\def\ep{\epsilon}
\def\varep{\varepsilon}
\newcommand{\tep}{{\tilde\epsilon}}
\newcommand{\tvarep}{{\tilde\varepsilon}}

\begin{document}
\papertitlepage
\setcounter{page}{0}

\preprintnumber{TIT/HEP-668}{UTHEP-718}

\vskip 7ex

\baselineskip=4ex
\begin{center} 
{\Large\bf\mathversion{bold}
MHV amplitudes at strong coupling  \\
\vskip 0.8ex
  and linearized TBA equations
}
\end{center}

\vskip 3ex 
\baselineskip=3.6ex
\begin{center}
 Katsushi Ito\footnote[1]{\tt ito@th.phys.titech.ac.jp},
  Yuji Satoh\footnote[2]{\tt ysatoh@het.ph.tsukuba.ac.jp} and
  Junji Suzuki \footnote[3]{\tt suzuki.junji@shizuoka.ac.jp}\\

\vskip 3ex
 
 \affiliation{1}
 {\it Department of Physics, Tokyo Institute of Technology}\\
 {\it Tokyo 152-8551, Japan}\\

\vskip 1ex
 \affiliation{2}
 {\it Institute of Physics, University of Tsukuba}\\
 {\it Ibaraki 305-8571, Japan}\\

\vskip 1ex
 \affiliation{3}
 {\it  Department of Physics, Shizuoka University}\\
 {\it Shizuoka 422-8529, Japan} 
 
\end{center}

\baselineskip=3.6ex
\vskip 4ex
\begin{center} {\bf Abstract} 
\end{center}
\vskip 1ex
The maximally helicity violating (MHV) amplitudes 
of ${\cal N} =4$ super Yang-Mills theory at strong coupling 
are obtained by solving
auxiliary thermodynamic Bethe ansatz (TBA) integral equations. 
We consider a limit where the TBA equations are linearized 
for large chemical potentials and masses therein.
By solving the linearized equations, we derive analytic expansions of the 6-point MHV amplitudes
in terms of the ratio of the chemical potential $A$ and the mass $M$.
The expansions are valid up to corrections exponentially small in $A$ or 
inversely proportional to powers of $A$.
The analytic expansions describe the amplitudes for small conformal cross-ratios 
of the particle momenta in a standard basis,
and  interpolate the amplitudes with equal cross-ratios and those in soft/collinear limits.
The leading power corrections are also obtained analytically.
We compare the 6-point rescaled remainder functions at strong coupling and at 2 loops
for the above kinematics.
They are rather different, in contrast to other kinematic regions discussed in the literature 
where they are found to be similar to each other.

\vskip 3ex
\vspace*{\fill}
\noindent
May, 2018
%
%

%%%
%\vspace*{5ex}
\setlength{\parskip}{1ex}
\renewcommand{\thefootnote}{\arabic{footnote}}
\setcounter{page}{0}
\newpage
%\baselineskip=3.5ex
%
%\tableofcontents
\setcounter{footnote}{0}
\setcounter{section}{0}
\pagestyle{plain}

\baselineskip=3.6ex
\newpage

\section{Introduction}

Developments in the study of the gauge-string duality and perturbative gauge theory
have merged into a deep understanding of the four-dimensional maximally supersymmetric 
Yang-Mills theory (${\cal N}=4$ SYM). It is now possible to address its dynamics 
even at finite coupling based on the underlying integrability in the planar limit \cite{Beisert:2010jr}:
one can find  
the spectrum of single-trace operators  
\cite{Gromov:2009tv,Bombardelli:2009ns,Arutyunov:2009ur}, and
a formulation has been given \cite{Basso:2013vsa}
to obtain the  scattering amplitudes, or equivalently 
\cite{Alday:2007hr,Drummond:2007aua,Brandhuber:2007yx,Drummond:2007cf,Drummond:2007au}, 
the expectation values 
of null-polygonal Wilson loops. 
Such integrability-based approaches have also been extended to the correlation functions
of single-trace operators \cite{Basso:2015zoa,Fleury:2016ykk}.

Focusing on the maximally helicity violating (MHV) amplitudes, 
the form of the $n$-point amplitudes is almost fixed 
by the Bern-Dixon-Smirnov (BDS) expression \cite{Bern:2005iz}
due to the anomalous conformal Ward identity \cite{Drummond:2007cf,Drummond:2007au}.
The evaluation of the full amplitudes thus reduces to finding its remainder 
which exists  for $n \geq 6$ beyond one loop \cite{Alday:2007he,Bern:2008ap,Drummond:2008aq}.
A notable fact to this end is that 
the analytic structure of the amplitudes is well controlled by the transcendentality and 
the associated symbol of loop integrals.  This enables us to ``bootstrap" the amplitudes.   
The complete and concise expression of 
the 6-point amplitudes  at 2 loops has been obtained in this way \cite{Goncharov:2010jf}, 
in agreement with the direct computation  \cite{DelDuca:2009au}.
The bootstrap method has been extended 
up to 5 loops for $n=6$ \cite{Dixon:2013eka,Dixon:2014voa,Caron-Huot:2016owq}, 
and up to 2 loops for $n=7$ \cite{Golden:2014xqf}.
For the restricted kinematics where the momenta of particles are two dimensional,
analytic results are also given in \cite{DelDuca:2010zp,Heslop:2010kq}.

At strong coupling, the MHV amplitudes are obtained \cite{Alday:2007hr} 
by evaluating the area of the minimal
surfaces in the five-dimensional anti-de Sitter space (AdS$_5$) whose boundary
ends on  the corresponding null-polygonal Wilson loops at the boundary of AdS$_5$.
Although only a few exact solutions are known for such null-polygonal minimal surfaces
\cite{Kruczenski:2002fb,Alday:2007hr,Sakai:2009ut,Sakai:2010eh,Basso:2014pla}, 
one can evaluate their area by solving 
auxiliary integral equations 
\cite{Alday:2009yn,Alday:2009dv,Alday:2010vh,Hatsuda:2010cc}.
These integral equations take the form of the thermodynamic Bethe ansatz (TBA) equations,
which appear in the  analysis of finite-size effects in integrable models. 

The equations for $n=6$ are indeed identified \cite{Alday:2009dv} with the TBA equations of 
the $\bbZ_4$-integrable model \cite{Koberle:1979sg}.
When the momenta of particles are restricted to three-dimensional spacetime,
the minimal surfaces are embedded in the AdS$_4$ subspace. In this case, 
the integral equations for the $n$-point amplitudes 
are identified \cite{Hatsuda:2010cc} with the 
TBA equations of the $su(n-4)_4/u(1)^{n-5}$ homogeneous sine-Gordon (HSG) 
model \cite{FernandezPousa:1996hi}.
For the two-dimensional kinematics for which the corresponding minimal surfaces 
are embedded in AdS$_3$, they are identified \cite{Hatsuda:2010cc} with
the  TBA equations of 
the $su(n/2-2)_2/u(1)^{n/2-3} $ HSG model.
The integrable models  in the general AdS$_5$ case
have not been identified yet for $n\geq 7$,  in particular due to the unusual 
nature of the integral equations.

The integral equations for the amplitudes can be solved numerically  by iteration.
In addition, the equations can be simplified by taking limits of the parameters therein.
For example, when the ``mass" parameters are large, the integrable model
reduces to a free massive theory in the infrared regime. The solution to the TBA equations is
then expressed by iterative  multiple integrals \cite{Zamolodchikov:1989cf}. 
 On the SYM side, the cross-ratios formed by particle momenta may become large or small
in this limit. 
This includes the collinear limit for which  the OPE/flux-tube expansions of the amplitudes
  are formulated at finite coupling \cite{Alday:2010ku,Basso:2013vsa}.
On the other hand, in the strict limit of vanishing masses (the ultraviolet regime),
the TBA equations are solved analytically \cite{Zamolodchikov:1989cf}.
From the point of view of SYM, it is the limit where the corresponding Wilson loops 
form regular polygons. Around this regular-polygonal limit, 
the amplitudes are expanded analytically by small masses
for $n=6$ in  the AdS$_5$ case  \cite{Hatsuda:2010vr,Hatsuda:2014vra}, and 
for general $n$ in the AdS$_4$  \cite{Hatsuda:2012pb}  and the AdS$_3$ 
\cite{Hatsuda:2011ke,Hatsuda:2011jn} case.

In this paper, we consider another limit of the TBA equations for the amplitudes,
where the ``chemical potentials" as well as the masses are large and hence
 the equations are linearized.  The corrections in the linearization are
  exponentially small in the chemical potentials, 
 or suppressed by their powers as understood by the Sommerfeld type argument \cite{JKStj95}.
In the case of the 6-point amplitudes, which we discuss concretely below, 
the linearized equations are solved following \cite{AlZamolodchikov1995}
as expansions  in the ratio of the mass $M$ and the chemical potential $A$ to any order. 
The power corrections are analyzed by 
extending the analysis in \cite{AlZamolodchikov1995}, and 
we show that the leading power corrections are analytically evaluated. 
The results are checked against the numerical solutions, to be found in agreement. 
These analyses of the corrections  assure that  the linearization
gives a controlled approximation of the original TBA equations for large $A$  and $M$.

Applying the solution of the linearized TBA equations,
we derive analytic expansions of the 6-point MHV amplitudes to any order.
As the parameters are varied, the cross ratios of the particle momenta in a standard basis 
are kept small and change from the equal value in the UV regime to those for 
the soft/collinear limits in the IR regime.
The amplitudes are well described by the expansion over the corresponding kinematic 
region from the UV regime to the IR regime.
Our results thus provide another concrete example where focusing on the strong coupling
enables us to explicitly evaluate the MHV amplitudes.
Since the collinear limit can be realized as the IR end point in the present case,
it is of  interest to consider implications of our expansion to the OPE/flux-tube expansion
at finite coupling.

Along the above trajectory of the cross-ratios, 
we also compare the 6-point remainder functions at strong coupling and at 2 loops
which are rescaled/normalized by their UV and IR values.
They turn out to be rather different
in contrast to the  cases where similarities are found between the strong-coupling results and
the perturbative results \cite{Brandhuber:2009da,Hatsuda:2011ke,Hatsuda:2011jn,Hatsuda:2012pb,Hatsuda:2014vra,Dixon:2013eka,Dixon:2014voa}.
This implies that the kinematic region described by our expansion  provides a 
 probe to study structural differences of the strong-coupling and the perturbative
results.

This paper is organized as follows: In section 2, we review the MHV amplitudes
at strong coupling and the associated TBA system. In section 3, we solve the
linearized TBA equations for the 6-point amplitudes, and analyze the 
corrections in the linearization inversely proportial to powers of $A$.
In section 4, we check the results in section 3
against numerical solutions. In section 5, 
we derive analytic expansions of the 6-point MHV amplitudes,
and compare the rescaled remainder functions at strong coupling 
and at 2 loops. We conclude with a summary and discussion in section 6.
Two appendices are also attached.  In appendix A, we evaluate 
one of the pseudo energies in a different way from the main text by direct integration.
This shows explicitly how the fractional powers of the spectral parameter  
appear from the summation over its integral powers,
in accordance with the periodicity which is required from the algebraic equations (Y-system)
associated with the TBA equations.
In appendix B, we estimate the derivative of a pseudo energy.

%%%
\section{Scattering amplitudes at strong coupling}
\label{sec:RevAmplitudes}

Let us consider 
the MHV amplitude ${\cal M}$  of ${\cal N} = 4$ SYM in the planar limit.
This is equivalent/dual to the expectation value of 
the null-polygonal Wilson loops whose edges correspond to the momenta of 
scattering particles 
\cite{Alday:2007hr,Drummond:2007aua,Brandhuber:2007yx,Drummond:2007cf,Drummond:2007au}. 
After the tree amplitude is factored out, the remaining scalar part at strong coupling 
is thus evaluated by the area ${\cal A}$ of the minimal surfaces in 
AdS$_5$ ending on  the null-polygonal Wilson loops at the boundary of AdS$_5$
\cite{Alday:2007hr}. Schematically,
\be
  {\cal M} \sim e^{-\frac{\sqrt{\lambda}}{2\pi} {\cal A}} \comma \nn
\ee
where $\lambda \gg 1$ is the 't Hooft coupling.
Since ${\cal N} =4$ SYM is conformal, natural kinematical variables to express
amplitudes are  the cross-ratios formed by  momenta of particles.

\subsection{TBA equations for MHV amplitudes}

Although only a few  solutions are known in closed forms
for such null-polygonal minimal surfaces
\cite{Kruczenski:2002fb,Alday:2007hr,Sakai:2009ut,Sakai:2010eh,Basso:2014pla}, 
one can evaluate their area by solving 
auxiliary integral equations 
\cite{Alday:2009yn,Alday:2009dv,Alday:2010vh,Hatsuda:2010cc}.
For  $n$-particle amplitudes, 
they take the form \cite{Alday:2010vh},
\eqb
\label{nptTBA}
\log Y_{2,s}(\theta) \Eqn{=} - \sqrt{2}  M_{s} \cosh\theta
-K_2 \ast {\cal L}_{1s} -K_1 \ast {\cal L}_{2s}  \comma \nn \\
\log Y_{1,s}(\theta) \Eqn{=} -M_{s}  \cosh\theta -C_s-\frac{1}{2}
K_2\ast {\cal L}_{2s} -K_1 \ast {\cal L}_{1s} - \frac{1}{2}K_3 \ast {\cal L}_{3s}  \comma \\
\log Y_{3,s}(\theta)  \Eqn{=} -M_{s}  \cosh\theta+C_s-\frac{1}{2}
K_2\ast {\cal L}_{2s} -K_1 \ast {\cal L}_{1s} + \frac{1}{2}K_3 \ast {\cal L}_{3s} \comma \nn
\eqe
where $M_s, C_s$ are constants, $s=1, ..., n-5$, and 
\eqb 
{\cal L}_{1s}
 \Eqn{=} \log \frac{(1+  Y_{1,s}) 
 (1+Y_{3,s}) }{(1+Y_{2,s-1})(1+Y_{2,s+1})} \comma \quad 
{\cal L}_{3s}
= \log \frac{ (1+ Y_{1,s-1})(1+Y_{3,s+1})}{ (1+  Y_{1,s+1})(1+ Y_{3,s-1}) } \comma  \nn \\
{\cal L}_{2s}
\Eqn{=}   \log \frac{(1+Y_{2,s})^2 }{(1+
Y_{1,s-1})(1+ Y_{1,s+1})(1+ Y_{3,s-1})(1+ Y_{3,s+1})} \period \nn
\eqe
 The Y-functions, $Y_{a,s}$,  are defined through the Stokes data 
of the auxiliary linear problem associated with the string equations of motion.
We have also denoted by $\ast$ the convolution 
$f \ast g (\theta) := \int_{-\infty}^\infty \frac{d\theta'}{2\pi} f(\theta-\theta')g(\theta')$,
with the kernels,
\be 
\label{kernels}
K_1(\theta) = \frac{1}{\cosh\theta} \comma  \quad 
K_2(\theta)=\frac{2\sqrt{2}\cosh\theta}{\cosh 2\theta} \comma\quad 
K_3(\theta)= 2i  \tanh 2\theta \period 
\ee
Numerically, these equations are solved by iteration 
where the initial values of 
$\log Y_{a,s}$  are  approximated 
by the ``driving terms", i.e. the terms not involving convolutions in  (\ref{nptTBA}).

Though $M_s$ are assumed  to be real and positive in (\ref{nptTBA}), 
they are complex  in general as $M_s = |M_s| e^{i\varphi_s} $. 
For small $\varphi_s$,  
the equations (\ref{nptTBA}) keep the same form but with
$ M_s \to |M_s| $, 
\begin{align}
\label{complexm}
 & 
  Y_{a,s}(\theta)\to  Y_{a,s}(\theta+i\varphi_s) \comma \quad
  K_{s,s'}^{a,a'}(\theta-\theta') \to K_{s,s'}^{a,a'}(\theta-\theta'+i\varphi_s-i\varphi_{s'})
 \comma
\end{align}
where   $K_{s,s'}^{a,a'}$  is the kernel $K_j$  for the 
the convolution involving  $Y_{a,s}$ and $Y_{a',s'}$.
When $ |\varphi_s - \varphi_{s+1}| $ exceeds $\pi/4$,
extra terms appear as the integrals pick up the poles of the integrands.

This formulation covers  signatures of the four-dimensional spacetime
other than the usual $(3,1)$ of $\bbR^{3,1}$. By  the reality condition
of the minimal surfaces, $C_s$ are required to be purely imaginary for the $(3,1)$
and $(1,3)$ signatures, whereas $C_s$ are real for the $(2,2)$ signature.
The number of  $|M_s|, \varphi_s, C_s$, i.e. $3(n-5)$, matches the number of 
the independent cross-ratios formed by the momenta of the scattering particles. 

The TBA equations (\ref{nptTBA}) can be converted to a set of algebraic equations,
called the  Y-system,
\be
\label{Ysystem}
  {Y_{a,s}^-Y_{4-a,s}^+ \over Y_{a+1,s}Y_{a-1,s}}
  = {(1+Y_{a,s+1})(1+Y_{4-a,s-1})\over(1+Y_{a+1,s})(1+Y_{a-1,s})}
  \comma
\ee
where $a=1,2,3$; $s=1, ..., n-5$; and $Y_{a,0} = Y_{a,n-4} = 0$ and $Y_{0,s} = Y_{4,s} = \infty$.
The superscripts $\pm$ stand for the shift of the argument, 
\be
  f^{\pm}(\theta) := f^{[\pm 1]}(\theta) \comma \qquad
  f^{[k]}(\theta) := f\Bigl( \theta + \frac{k}{4}\pi i\Bigr) \period \nn
\ee
From this Y-system,  the Y-functions turn out to have the (quasi-)periodicity
\cite{AlZamolodchikov1991,FominZelevinsky2007,IIKKN1,IIKKN2},
\be
\label{Yperiod}
   Y_{a,s}^{[n]} (\theta) = \left\{ 
   \begin{array}{ll}
     Y_{a, n-4-s}(\theta)  &  (s : {\rm odd}) \\
     Y_{4-a, n-4-s}(\theta)  & (s : {\rm even})
   \end{array}
  \right.
  \period
\ee
The Y-system can also be used to obtain the Y-functions with large imaginary shift 
of the argument, for which  the TBA equations (\ref{nptTBA})
are modified due to the pole contributions.
For a review of Y-systems, see for example \cite{Kuniba:2010ir}.

The integral equations of the type (\ref{nptTBA}) appear 
in the analysis of finite-size effects in integrable models,
and are called the thermodynamic Bethe ansatz 
equations \cite{Zamolodchikov:1989cf}.
As mentioned in the introduction,
the equations (\ref{nptTBA}) are identified \cite{Alday:2009dv} with the TBA equations of 
the $\bbZ_4$-integrable model \cite{Koberle:1979sg} for $n=6$.
When $Y_{1,s} = Y_{3,s}$ and hence  $C_s = 0$, 
the minimal surfaces are embedded 
in AdS$_4$, which describe the scattering of the special 
kinematics with three-dimensional momenta.
In this case, (\ref{nptTBA}) are identified \cite{Hatsuda:2010cc} with the 
TBA equations of the $su(n-4)_4/u(1)^{n-5}$ homogeneous sine-Gordon 
model \cite{FernandezPousa:1996hi}.
By imposing further constraints, the minimal surfaces are embedded in AdS$_3$, 
which describe the scattering for the two-dimensional kinematics. In this case,
$n$ is even because of the momentum conservation, and 
(\ref{nptTBA}) become the  TBA equations of 
the $su(n/2-2)_2/u(1)^{n/2-3} $ HSG model \cite{Hatsuda:2010cc}.
In the integrable models, the logarithms  of the Y-functions $Y_{a,s}$ are  pseudo energies,
$M_s$  are masses of particles (measured in the unit of the inverse system size $1/L)$, 
and $C_s$ are chemical potentials. 
The corresponding integrable models  in the AdS$_5$ case
have not been identified yet for $n\geq 7$.
From the TBA point of view, 
what is unusual in (\ref{nptTBA}) in this case is that the kernel $K_3(\theta)$ does not
decay for large $|\theta|$.
\subsection{Area, remainder function and cross-ratios}

After a proper regularization, the area is expressed for 
 $ n  \notin 4 \bbZ$ as
\be
\label{regArea}
  {\cal A} =  A_{\rm div} + A_{\rm BDS-like} + A_{\rm periods} + A_{\rm free} 
  \comma
\ee
up to a constant.
Here, $A_{\rm div} $ is a divergent term, and $A_{\rm BDS-like}$ is the term
which satisfies the anomalous conformal Ward identity \cite{Drummond:2007cf,Drummond:2007au}, 
as the Bern-Dixon-Smirnov (BDS) expression \cite{Bern:2005iz}. 
The third term $ A_{\rm periods}$ comes from period integrals
associated with the underlying auxiliary linear problem or the Hitchin system, and is
 expressed by  the mass parameters 
$M_s = |M_s| e^{i\varphi_s}$.
The explicit forms of   $A_{\rm div}, A_{\rm BDS-like}$ and $A_{\rm periods} $
are found in \cite{Alday:2009yn,Alday:2009dv,Alday:2010vh}. 
The last term $A_{\rm free}$ is obtained from  
the solution to the TBA equations (\ref{nptTBA}).
It coincides with the free energy of the corresponding integrable model 
(when it exits). Explicitly,
\be
A_{\rm free}  =  \sum_{s}  \int_{-\infty}^\infty { d \theta \over 2 \pi } | M_s|\cosh \theta \cdot
    \log \left[ \bigl(1 + {Y}_{1,s}(\theta_s)\bigr)
        \bigl(1 + {Y}_{3,s}(\theta_s)\bigr) \bigl(1 + {Y}_{2,s}(\theta_s)\bigr)^{\sqrt{2} } \right]
        \comma \nn
\ee        
with $\theta_s := \theta + i \varphi_s$.
For $n \in 4 \bbZ$,  the expression of the area may be obtained by taking an appropriate
limit from  $n \notin 4 \bbZ$.
Given the formula (\ref{regArea}), the non-trivial part of the area/strong-coupling amplitudes
reduces to $A_{\rm free}$. 

Since the structure of the MHV amplitudes is almost captured by the BDS expression, 
it may be sufficient to consider its remainder \cite{Alday:2007he,Bern:2008ap,Drummond:2008aq},
\be
\label{remainder}
  R := -({\cal A} - A_{\rm div} - A_{\rm BDS}) =  \Delta A_{\rm BDS}  - A_{\rm periods} - A_{\rm free} 
  \comma
\ee
where $A_{\rm BDS}$ is the (finite part of the) BDS expression, and 
$ \Delta A_{\rm BDS}:= A_{\rm BDS} - A_{\rm BDS-like}$.  
This remainder functions is conformally invariant and is a function of the cross-ratios
 of the particle momenta $p^\mu_j$,
\be
  \chi_{ijkl} := \frac{x^2_{ij} x^2_{kl}}{x^2_{ik} x^2_{jl}} \comma \quad x^\mu_{ij} := x^\mu_i - x^\mu_j \comma
  \quad p^\mu_j = x_j^\mu - x_{j+1}^\mu \period \nn
\ee
The momenta form null polygons corresponding to 
the dual Wilson loops due to the momentum conservation.
The subscript of the cusp points $x_j$ is hence understood modulo $n$. 
These cross-ratios are expressed by 
the Y-functions at special values of the argument, e.g.
\be
   \label{chiijkl}
   \chi_{k,-k,-k-1,k-1} = U_{2k-2}^{\langle 0 \rangle} \comma
  \quad 
  \chi_{k+1,-k,-k-1,k} = U_{2k-1}^{\langle 1 \rangle} \comma \nn
\ee
where we have defined
\be
 \label{UY}
  U_s(\theta) := 1+ \frac{1}{Y_{2,s}(\theta)} \comma
  \quad f^{\langle k \rangle} := f^{[k]}(0) = f\Bigl( \frac{k}{4}\pi i \Bigr) \period
\ee
A shift of  $\theta$ 
 induces a cyclic shift of the cusp points,
\be
  \label{Uchi}
   U_{2k-2}^{\langle 2r \rangle} = \chi_{k+r,-k+r,-k-1+r,k-1+r} 
   \comma \quad
    U_{2k-1}^{\langle 2r +1 \rangle} = \chi_{k+1+r,-k+r,-k-1+r,k+r}  \period
\ee
 The product of these cross-ratios yields generic ones.

A useful parametrization of the cross-ratios is 
given by the coordinates $(\tau_s, \sigma_s, \phi_s)$ 
associated with the symmetries of (parts of) the null-polygons \cite{Alday:2010ku,Sever:2011pc}.
They are directly related to the Y-functions,
\be
\label{hatYcr}
  \widehat{Y}_{1,s}(0) = e^{\phi_s - \sigma_s - \tau_s} \comma \qquad
  \widehat{Y}_{2,s}(0) = e^{-2 \tau_s} \comma \qquad
  \widehat{Y}_{3,s}(0) = e^{-\phi_s - \sigma_s - \tau_s} \comma
\ee
where $s=1, ..., n-5$, and 
\be
    \widehat{Y}_{a,s} (\theta) := \left\{ 
     \begin{array}{ll}
        Y_{a,s}   (\theta)  &   (a+s : {\rm even}) \\
        Y^-_{a,s}  (\theta)   &  (a+s : {\rm odd})
     \end{array}
    \right.
    \period \nn
\ee
In the OPE approach \cite{Alday:2010ku,Basso:2013vsa}, 
these are used to compute the finite coupling amplitudes, which are not restricted to
the MHV case. The amplitudes there are expanded by the contributions 
from the flux-tube excitations around the multi-collinear limit $\tau_s \to \infty$. 
Re-summing  over such series recovers the TBA equations at strong coupling 
\cite{Fioravanti:2015dma,Bonini:2015lfr}.
The explicit relation of  $(\tau_s, \sigma_s, \phi_s)$ and the cross-ratios $\chi_{ijkl}$
in (\ref{chiijkl}) is found by using (\ref{UY}), (\ref{Uchi}) and the Y-system (\ref{Ysystem}).

%%%
\subsection{Limits of the TBA system}
The analysis of the TBA equations (\ref{nptTBA}) can be simplified
by taking the limits of the parameters therein.

\par\smallskip
As for the mass parameters $|M_s|$, 
there are two simple limits 
where  $Y_{a,s}$ and $A_{\rm free}$ are analytically evaluated.

\ni
{\bf Large mass/IR limit :}
One is the limit where all  $|M_s| \gg 1$. This is the low temperature/IR limit
where the integrable models  reduce to free massive theories.
The Y-functions are given by the driving terms, and the free energy trivially vanishes.
From the point of view of SYM, it is the limit where cross-ratios can be large or small.
Around this limit, the Y-functions and the free energy are expanded 
by iterative multiple integrals \cite{Zamolodchikov:1989cf}.
By analytic continuation, this limit is also connected to the Regge limit \cite{Bartels:2010ej}.
Further adjusting the phases $\varphi_s$ yields
the above-mentioned multi-collinear limit.

\par\smallskip
\ni
{\bf Small mass/UV limit :}
The other  is the limit where all $|M_s| \ll 1$. 
This is the high temperature/UV limit
where the integrable models reduce to conformal field theories (CFTs). 
From the point of view of SYM, it is the limit where the dual 
null polygonal Wilson loops become $\bbZ_n$-symmetric 
(regular-polygonal).
By the standard method on the TBA system \cite{Zamolodchikov:1989cf},
or from the Y-system (\ref{Ysystem}) with the $\theta$-dependence dropped,
the Y-functions and the free energy are evaluated explicitly 
in the strict limit of vanishing masses.
We supply a concrete example for $n=6$ shortly in subsection \ref{6pt}.
Around this limit, the  free energy is expanded by conformal 
perturbation \cite{Zamolodchikov:1989cf}. 

For the hexagonal minimal surfaces in AdS$_5$, the Y-functions and hence the amplitudes
are expanded around the small mass limit based on the quantum  Wronskian 
relation
\cite{Hatsuda:2014vra} following 
an earlier work \cite{Hatsuda:2010vr}.
For the minimal surfaces in  AdS$_3$ \cite{Hatsuda:2011ke,Hatsuda:2011jn}
and AdS$_4$ \cite{Hatsuda:2012pb}, 
the small-mass expansions of the Y-functions and the amplitudes are derived
based 
on the relation \cite{Bazhanov:1994ft,Dorey:1999cj,Dorey:2005ak} 
between the $g$- and Y-functions using auxiliary 
boundary conformal perturbation. 
The expansions by the conformal perturbation are in terms of  the couplings 
of the relevant operators to the CFT in the UV limit. They are expressed by 
the mass (and other) parameters of the TBA system via the mass-coupling relation.
For the hexagonal minimal surfaces in AdS$_5$,
the corresponding $\bbZ_4$ integrable model has 
a single mass parameter, and
the relevant exact mass-coupling relation is given in \cite{Fateev:1993av} based on the work
\cite{AlZamolodchikov1995}. 
When the models have multi-scales, it is in general
difficult to obtain such relations.  In \cite{Bajnok:2015eng,Bajnok:2016ocb}, 
for the analysis of multi-scale integrable models, 
the exact mass-coupling relation is obtained in the $n=10$ case
for the AdS$_3$ minimal surfaces. (For AdS$_3$, the TBA system has 
multi-scales for $n \geq 10$.) 

\par\smallskip
\ni
{\bf Phases :} 
As for the phases $\varphi_s$,  they appear as the imaginary shifts 
of $Y_{a,s}(\theta)$
as in (\ref{complexm}). Their change
induces the ``wall-crossing" phenomenon, or the change of the
form of the TBA equations (\ref{nptTBA}). They are 
restricted to a  finite range because of the periodicity of the Y-functions (\ref{Yperiod}).

\par\smallskip
\ni
{\bf Large chemical potentials :} 
As for the last parameters $C_s$,  it follows from the TBA equations  
that 
\be
 \log \frac{Y_{3,s}}{Y_{1,s}} = 2 C_s + K_3 \ast {\cal L}_{3s} \sim 2 C_s  \comma \nn
\ee
where we have used a rough estimate for large $C_s $ and $ |M_s| $ such that
$\log Y_{a,s} $ are approximated in (\ref{nptTBA}) by the driving terms. 
From (\ref{hatYcr}), one  finds that $C_s$ are directly related 
to $\phi_s$ (Lorentz boost coordinates) in the cross-ratios.
In the following, we consider the case  of $n=6$ and demonstrate 
that the large potentials $C_s$ indeed provide
another useful limit 
in analyzing the TBA system for the amplitudes.
Precisely, we combine two limits, large mass and chemical potential, 
which has not yet been considered so far.

\subsection{6-particle amplitudes}
\label{6pt}

When the number of the particles $n=6$, there are only three non-trivial Y-functions 
 $Y_{1,1}, Y_{2,1}, Y_{3,1}$. Thus, ${\cal L}_{3s}$ 
in (\ref{nptTBA}) is vanishing, and the ratio $Y_{3,1}/Y_{1,1}$ is a constant.
The TBA equations then reduce to 
\eqb
 \label{TBA6}
   \varep(\theta) \Eqn{=}
     -A + M \cosh \theta + K_1 \ast \log(1+\mu^{-2} e^{-\varep})(1+e^{-\varep}) 
    +  K_2 \ast \log(1+e^{-\tvarep}) \comma \nn \\
    \tvarep(\theta) \Eqn{=}  
     \sqrt{2} M \cosh \theta + K_2 \ast \log(1+\mu^{-2} e^{-\varep})(1+e^{-\varep}) 
    + 2 K_1 \ast \log(1+e^{-\tvarep}) \comma
\eqe
where we have set  
\be
\label{Yep}
   \log Y_1(\theta+i\varphi) = - \varep(\theta) - 2A \comma \quad  
   \log  Y_3(\theta+i\varphi) =: - \varep(\theta) \comma \quad 
    \log Y_2(\theta+i\varphi) =: - \tvarep(\theta) \period
\ee
with
\be
  Y_a(\theta) := Y_{a,1}(\theta)   \comma \quad 
     M :=  | M_1 |
   \comma \quad \mu := e^A \comma \quad A := C_1 \comma
   \quad \varphi:= \varphi_1 \period \nn
\ee
From now on, we assume $|\varphi| < \pi/4$, 
unless otherwise stated,  so that the TBA equations 
take the form as in (\ref{nptTBA}) without the pole contributions.
The Y-system (\ref{Ysystem}) for $n=6$ reads
\be
 \label{Ysystem6}
   Y_1^+ Y_3^- = \frac{Y_2}{1+Y_2} \comma \quad 
   Y_2^+ Y_2^- = \frac{Y_1 Y_3}{(1+Y_1)(1+Y_3)} \period
\ee
Then the equation for $Y_3^+ Y_1^-$ is equivalent to the  first  one 
because $Y_3/Y_1 = \mu^2$. From (\ref{Yperiod}) or  (\ref{Ysystem6}), 
it follows that 
\be
\label{period6}
Y_a^{[6]} = Y_a \period
\ee
The components of the area or the remainder function in (\ref{remainder}) are
\be
\label{Afree}
  A_{\rm free} = \int_{-\infty}^\infty \frac{d\theta}{2\pi} 
 M \cosh \theta \times \log(1+\mu^{-2} e^{-\varep(\theta)}) (1+e^{-\varep(\theta)})
 (1+ e^{-\tvarep(\theta)})^{\sqrt{2}} \comma
\ee
and
\be
\label{AperiodBDS6}
   A_{\rm period} = \frac{1}{4} M^2 \comma \qquad
   \Delta A_{\rm BDS} = - \frac{1}{4} \sum_{j=1}^3 \Li_2\Bigl( 1- \frac{1}{u_j}\Bigr)
   \period
\ee

For the 6-point amplitudes, there are three independent cross-ratios.
A standard basis for them  is
\be
\label{ujUj}
  u_1 := \chi_{2653} = {1 \o U^{\langle 3 \rangle} }\comma \quad
  u_2 := \chi_{3164} = {1 \o U^{\langle 5 \rangle} } \comma \quad
  u_3 := \chi_{4215} = {1 \o U^{\langle 1 \rangle} }\comma
\ee
with $U:= U_1$.
They are also parametrized by the variables of the type (\ref{hatYcr}),
\be
\label{Ytsp}
  Y_1^{\langle 0 \rangle} = e^{\phi - \sigma - \tau} \comma \quad 
  Y_2^{\langle -1 \rangle} = e^{- 2\tau} \comma \quad
  Y_3^{\langle 0 \rangle} = e^{-\phi - \sigma - \tau} \comma
\ee
with $(\tau,\sigma, \phi) := (\tau_1, \sigma_1, \phi_1)$. 
From the periodicity (\ref{period6}),  one has $Y_2^{\langle -1 \rangle} = Y_2^{\langle 5 \rangle} $, 
and $u_2$ is readily related to  $\tau$. 
Furthermore, $u_{1,3}$ are related to $(\tau,\sigma, \phi) $ through
\be
 Y_2^{\langle 1 \rangle} 
 = \frac{Y_1^{\langle 0 \rangle}Y_3^{\langle 0 \rangle}}{
  (1+Y_1^{\langle 0 \rangle})(1+Y_3^{\langle 0 \rangle})Y_2^{\langle -1 \rangle}} 
  \comma \quad
 Y_2^{\langle 3 \rangle} 
  = \frac{ Y_2^{\langle -1 \rangle} }{ 
  \bigl(Y_2^{\langle -1 \rangle}+Y_1^{\langle 0 \rangle}(1+Y_2^{\langle -1 \rangle}) \bigr)
 \bigl( Y_2^{\langle -1 \rangle}+Y_3^{\langle 0 \rangle}(1+Y_2^{\langle -1 \rangle}) \bigr)}
 \comma \nn
\ee
which follows from the Y-system.
We thus find 
\cite{Basso:2013vsa},%
\footnote{
Comparing with the parametrization in \cite{Alday:2010ku},
$e^{-\tau_{\rm there}} =e^{\tau}(-1+\sqrt{1+e^{-2\tau}})$, 
$  e^{-\sigma_{\rm there}} = e^{-\sigma} \sqrt{1+e^{-2\tau}}$.
}
\be
\label{utsp}
  \frac{1}{u_2}  {=}  1+e^{2\tau} \comma \quad 
    \frac{u_1}{u_2u_3} {=}  e^{2\sigma+2\tau} \comma \quad
     \frac{1}{u_3} = 1+e^{-2\tau} (1+e^{\tau+\sigma+\phi})(1+e^{\tau+\sigma-\phi}) \period
\ee

\par\smallskip
\ni
{\bf UV and IR limits :} 
For reference for the discussion below, 
we summarize the behavior of the pseudo energies, the free energy and the remainder
function in the UV and the IR limit.
First, in the UV limit $M \to 0$ with $A$ and $\varphi$ fixed, 
\begin{align}
\label{smallM}
  &e^{\varep} \to e^{\varep_{\rm UV}} := \frac{1}{\mu} (\mu^{1/3} + \mu^{-1/3})
  \comma \quad 
  e^{\tvarep} \to  e^{\tvarep_{\rm UV}} :=1+ \mu^{2/3} + \mu^{-2/3} \comma \nn \\
 &A_{\rm free} \to  A_{\rm free, UV} := \frac{\pi}{6} c + \frac{A^2}{3\pi}
 \comma 
\end{align}
where $c=1$ is 
the central charge of the CFT in the UV limit of the underlying $\bbZ_4$ integrable model. 
These imply all $u_j \to \hat{u}_{\rm UV} := 1/(\mu^{1/3} + \mu^{-1/3})^{2}$, and 
\be
\label{RUV}
 R \to R_{\rm 6, UV} := 
 - \frac{3}{4} \Li_2\Bigl(1- \frac{1}{\hat{u}_{\rm UV}}\Bigr) - \frac{\pi}{6}  -  \frac{A^2}{3\pi}  \period
\ee 
As mentioned above, the boundary of the minimal surface 
becomes regular-polygonal in this limit.

On the other hand, in the IR limit $M \to \infty $ with $A$ and $\varphi$ fixed,
\begin{align}
\label{largeM}
 & \varep \to   \varep_{\rm IR} := -A + M \cosh \theta \comma \quad
  \tvarep \to \tvarep_{\rm IR} := \sqrt{2} M \cosh \theta \comma \nn \\
 & A_{\rm free} \to  A_{\rm free,IR} :=0 \period
\end{align}
Again from the Y-system, 
these imply that two of $u_j \to 0$ and  the other $u_k \to 1$ for generic $\varphi$,
and
\be
\label{RIR}
 R \to  R_{\rm 6, IR} := \frac{\pi^2}{12} \period
\ee
This is the soft limit, e.g. where $u_1 \sim1, u_{2,3} \sim 0$ and both $\tau, \sigma \gg 1$
 for  $-\pi/4 < \varphi < 0$ \cite{Alday:2009dv}.
Sending $\varphi \to -\pi/4$, one can achieve the collinear limit where 
$u_2 \sim 0$, $u_1+u_3 \sim 1$ and 
$\tau \gg 1$ with $\sigma$ fixed. This is the regime of  the OPE expansion 
\cite{Alday:2010ku,Basso:2013vsa}.
By changing $\varphi$, the roles of $u_j$ are permuted.
For higher order terms in the UV and the IR limit, we refer to 
\cite{Hatsuda:2014vra,Hatsuda:2010vr}.

\begin{figure}[t]
\vspace{5ex}
 \begin{center}
   \begin{minipage}{0.4\hsize}
   \begin{center}
  \includegraphics[width=55mm]{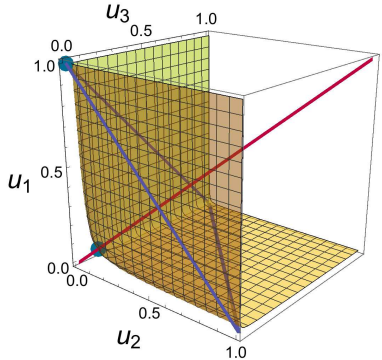} 
 \end{center}
 \end{minipage}
\caption{Cross-ratios and various limits.
The red straight line corresponds to the UV limit. 
The  triangle with vertices $(u_1, u_2, u_3) = (1,0,0), (0,1,0), (0,0,1)$
corresponds to the IR limit. In particular,
its sides and vertices correspond to the collinear and the soft limit,
respectively.}
 \label{fig:Cross-ratios}
\end{center}
\end{figure}

Figure \ref{fig:Cross-ratios}
 illustrates these limits in the space of the three cross-ratios $(u_1, u_2, u_3)$.
The red straight line represents $u_1 = u_2 = u_3$ corresponding to the UV limit.
The   triangle represents the IR limit.
Its sides correspond to the collinear limits, whereas its vertices 
$(u_1, u_2, u_3) = (1,0,0), (0,1,0), (0,0,1)$
to the soft limit. 
The yellow surface represents $u_1u_2u_3 = e^{-2A}$, which is discussed 
in section \ref{sec:RExpansion}.
Based on the linearized TBA equations in the next section,
the expansions of $Y_a, A_{\rm free}$ and $R$ for large $A$ and $M$ are derived there.
They interpolate the kinematic points in the UV and the IR limit, 
e.g. which are marked by the green blobs 
for generic $-\pi/4 < \varphi < 0$. When the phase is adjusted, e.g. as  $\varphi \to -\pi/4$,
the point in the IR limit moves onto a side of the triangle away from its vertices.

%%%%
%%%%

%%%
\section{Linearized TBA equation}
\label{sec:LTBA}

In this section, we consider the TBA equations for the 6-point amplitudes (\ref{TBA6})
for large $A$ and $M$ where the equations  are linearized.
We first introduce the linearlized TBA equations in subsection \ref{sec:Linear} and 
summarize the solution to the linearized equations following \cite{AlZamolodchikov1995}
in subsection \ref{sec:LTBA1}. 
There are several types of the corrections in the linearization.
One is exponentially small in $A$ (as in (\ref{ExSmall})), another is expected to be
exponentially small in $A$ (as in $\Delta_2$ in (\ref{Delta12})), and the other
is suppressed by  the powers of $A$ (as in $\Delta_1$ in (\ref{Delta12})) 
due to the Sommerfeld type argument \cite{JKStj95}.
The exponential behavior  of the second type is confirmed 
by numerically checking the scaling of the pseudo energies 
in the next section. As for the third type, the power series
corrections are analyzed by 
extending the analysis in \cite{AlZamolodchikov1995}.
 We  explicitly solve the linearized equations including 
 the leading term in the power series
 corrections in subsection \ref{sec:LTBA2}.
The free energy with the first $\calO(L^{-2})$  corrections
is given in subsection \ref{sec:LTBAf}. 
In subsection \ref{sec:MAexp},
the leading terms are re-expanded 
in terms of the parameters in the TBA equations, i.e. $M/A$.
These results are applied to the amplitudes in section \ref{sec:RExpansion}.
The analysis of the corrections assures that  the linearization
gives a controlled approximation which is 
valid up to the relative corrections which are exponentially small in $A$ or of order $1/A^2$.

\subsection{Linearization}
\label{sec:Linear}
We consider the limit, 
\begin{equation}\label{eq:the_limit}
A= aL, \qquad M=mL,\qquad  a,m ={\cal O}(1), \quad  L \gg 1.
\end{equation}
The TBA equations  (\ref{TBA6})
imply the following representation of  $\varepsilon(\theta)$ and $\tilde{\varepsilon}(\theta) $, 
\[
\varepsilon(\theta) =L( -a+m \cosh\theta) +r_1(\theta ),  \qquad 
\tilde{\varepsilon}(\theta) = L \sqrt{2} m \cosh\theta +r_2(\theta ), 
\] 
where $r_i(\theta )$ are positive numbers. 
Hence some terms in (\ref{TBA6}) 
are exponentially small in $L$ for $\theta \in \mathbb{R}$:
\begin{align} 
\label{ExSmall}
\log(1+\mu^{-2} {\rm e}^{-\varepsilon} )  & \sim {\rm e}^{-L(a+m \cosh \theta +r_1)} 
< {\rm e}^{-L(a+m \cosh \theta)} , \nn \\
 \log(1+{\rm e}^{-\tilde{\varepsilon}})  & \sim {\rm e}^{-L( \sqrt{2} m  \cosh \theta +r_2)} 
 <  {\rm e}^{-L \sqrt{2} m  \cosh \theta }. 
\end{align}
By neglecting these terms,   we approximate the first equation in (\ref{TBA6}) 
 by the form containing only ${\varepsilon}$,
\begin{equation}\label{eq:tba1d}
\varepsilon(\theta) = -A+M \cosh\theta +K_1 * \log(1+{\rm e}^{-\varepsilon} ) .
\end{equation}
The other function $\tilde{\varepsilon}$ is then evaluated 
by using ${\varepsilon}$ as
\begin{equation}
\tilde{\varepsilon}(\theta) =\sqrt{2} M \cosh\theta +K_2 * \log(1+{\rm e}^{-\varepsilon} ) .
\nn
\end{equation}

We assume that there are unique Fermi points $\theta = \pm B$ satisfying 
\[
\varepsilon( \pm B) =0.
\]
The explicit expression of $B$
in terms of $A$ and $M$ will be determined a posteriori.
The convolution term is  then divided into three pieces, 
\begin{equation}\label{eq:convolution3} 
K_1 * \log(1+{\rm e}^{-\varepsilon} )  
= -\int_{-B}^B K_1(\theta-\theta') \varepsilon(\theta') \frac{d\theta'}{2\pi} +\Delta_1(\theta)+ \Delta_2(\theta) \comma
\end{equation}
where
\begin{align} 
\label{Delta12}
\Delta_1(\theta)&= \Bigl( \int_{|\theta'-B|<\eta} + \int_{|\theta'+B|<\eta}\Bigr)
 K_1(\theta-\theta') \log(1+{\rm e}^{-|\varepsilon(\theta') |}) \frac{d\theta'}{2\pi}  \comma \nn \\
\Delta_2(\theta)&= \Bigl( \int_{\theta'>B+\eta} + \int_{\theta'<-B-\eta}+ \int_{|\theta'|<B-\eta}\Bigr)
 K_1(\theta-\theta') \log(1+{\rm e}^{-|\varepsilon(\theta') |}) \frac{d\theta'}{2\pi} \comma
\end{align} 
and $\eta =O(1)$.
 By keeping only the first term in the right hand side of (\ref{eq:convolution3}),
 one obtains a linearized TBA equation from (\ref{eq:tba1d}),
 \begin{equation} \label{eq:tba1dd} 
 \varepsilon(\theta) = -A+M \cosh\theta-\int_{-B}^B K_1(\theta-\theta') \varepsilon(\theta') \frac{d\theta'}{2\pi}.
 \end{equation}
 Similarly  we have
  \begin{equation} \label{eq:tba2dd} 
 \tilde{\varepsilon}(\theta) =\sqrt{2}M \cosh\theta-\int_{-B}^B K_2(\theta-\theta') \varepsilon(\theta') \frac{d\theta'}{2\pi}.
 \end{equation}
The eq. (\ref{eq:tba1dd})  is identical to the $p=\frac{1}{2}$ case of (3.3)  in \cite{AlZamolodchikov1995}
 by 
 \[
 \tilde{K}(\theta)=\delta(\theta) +\frac{K_1(\theta)}{2\pi}, \qquad
 \epsilon_{{\rm Zamolodchikov }}(\theta)= -\varepsilon(\theta). 
 \]
We can thus follow  \cite{AlZamolodchikov1995} to treat it analytically.
 
In the limit (\ref{eq:the_limit}), we expect from the linearized equations
(\ref{eq:tba1dd}), (\ref{eq:tba2dd}) that
\be
  \label{eq:epscale}
  \varepsilon(\theta) \sim L \epsilon(\theta),  \qquad  \tilde{\varepsilon}(\theta) 
   \sim L \tilde{\epsilon}(\theta)
  \qquad    \epsilon, \tilde{\epsilon} \sim O(1).
\ee
Under the assumption of this scaling,
$\Delta_2$ is exponentially small in $L$.
The scaling  will be confirmed numerically in the next section.
Below we are  interested in the corrections to the linearized equations
due to $\Delta_1(\theta)$. 
The Sommerfeld type argument results in, as the simplest  approximation \cite{JKStj95},
\begin{equation}\label{eq:sommerfeld}
\Delta_1(\theta) \sim \frac{\pi}{12 |\varepsilon'(B)|} (K_1(\theta-B)+ K_1(\theta+B)).
\end{equation}
 This turns out to yield an $\calO(L^{-2})$ correction to
$\epsilon(\theta)$.
One can  systematically analyze the higher power terms in  $L^{-2}$ in principle
by incorporating higher order terms from $\Delta_1$.
We, however, restrict ourselves to the leading correction below.
 
%%%
\subsection{Analysis of the linearized TBA }
\label{sec:LTBA1}
We recall the results in \cite{AlZamolodchikov1995} 
which are necessary  in the following discussion.
Let us introduce
\[
\epsilon_B(\theta)=
\begin{cases}
-\frac{\varepsilon(\theta)}{L}= -\epsilon(\theta)&  |\theta|<B,\\
0&    |\theta| \ge B.\\
\end{cases}
\]
Then the linearized TBA equation (\ref{eq:tba1dd}) is equivalent to,
\begin{equation} \label{linearALz} 
D_{\epsilon}(\theta) = \int_{-\infty}^{\infty} \tilde{K}(\theta-\theta') \epsilon_B(\theta').
\end{equation}
The function $D_{\epsilon}(\theta)$ in the left hand side is defined by 
\begin{equation}
D_{\epsilon}(\theta)=Y(\theta)+Y(-\theta), 
\qquad
Y(\theta)= 
\begin{cases}
X(\theta)&   \theta>B  , \\
\frac{a- m{\rm e}^{\theta}}{2}&  \theta<B. 
\end{cases}
\nn
\end{equation}
Here $X(\theta)$ is an unknown function of $\theta$.

The (inverse) Fourier transformation for  an arbitrary smooth function $f(\theta)$  is defined by
$$
\widehat{f}(w) =\int_{-\infty}^{\infty} f(\theta) {\rm e}^{iw\theta} d\theta ,
\qquad
f(\theta)= \int_{-\infty}^{\infty} \widehat{f}(w) {\rm e}^{-iw\theta}  \frac{dw}{2\pi}.
$$
The kernel $\tilde{K}$ admits the factorization in the Fourier space,
\begin{align*}
\widehat{K}(w) &=\frac{1}{\Kp(\omega) \Km(\omega)},   \qquad  \Km(w)=\Kp(-w), 
\end{align*}
where $\Kp (\Km) $ is analytic in the lower (upper) half plane
 and  $\widehat{K}_{\pm}(w) \rightarrow 1$ as $|w| \rightarrow \infty$.
The explicit form of $\Kp(w)$ reads,
\[
\Kp(w)= \sqrt{\frac{2\pi}{3}} {\rm e}^{ i\Delta w}  
\frac{\Gamma(1+\frac{3 w}{4}i)}{\Gamma(1+\frac{w}{4}i) \Gamma(\frac{1}{2}+\frac{w}{2}i) },
\qquad
\Delta=-( \frac{3}{4} \log\frac{3}{2}+\frac{1}{4}\log 2 ).
\]

We further introduce $v(w)$  by
\[
v(w) = {\rm e}^{-i w B} \Km(w) Y(w).
\]
Based on the Wiener-Hopf type factorization, it is shown  in \cite{AlZamolodchikov1995}
that  eq. (\ref{linearALz})  is casted into   the integral equation  for $v(w)$,
\begin{align}
\label{integeqv}
v(w) &=D_v(w)+\int_{-\infty}^{\infty} \frac{ {\rm e}^{2iw'B}}{w+w'+i0^+} 
\alpha(w') v(w')  \frac{dw'}{2\pi i} ,
\nn \\
&D_v(w)= -i\frac{a \Kp(0)}{2(w-i 0^+)} + i \frac{m e^{B} \Kp(-i)}{2(w-i)} ,
 \\
&\alpha(w)= \frac{\Kp(w)}{\Km(w)}.  \nonumber
\end{align}

The integral equation  (\ref{integeqv})  concludes that $v(w)$ 
is an ``almost" pole free function in the upper half plane and takes the form,
\begin{align}
v(w)&=  -i\frac{a \Kp(0)}{2}(\frac{1}{w-i0^+}+\frac{1}{w+i0^+})   + i \frac{m e^{B} \Kp(-i)}{2(w-i)}
+ \sum_{n\ge 1} \frac{ {\rm e}^{-\frac{8}{3} n B}}{w+\frac{4}{3}n i} \alpha_n v_n,  \nn \\
\alpha_n &= {\rm res}_{w=\frac{4}{3}n i} \alpha(w),  \qquad
v_n =  v(\frac{4}{3}n i).   \nonumber 
\end{align}
This  determines the coefficients $v_n$  in a self-consistent way. 
In order to determine them,
we utilize the condition $\varepsilon(B)=0$.
This implies $v(w) \sim O(1/w^2)$ for $|w| \gg 1$, which is 
written as 
\begin{equation} \label{epsilonBeq0}
i a \Kp(0)-\frac{im {\rm e}^{B}}{2} \Kp(-i) = 
\int_{C_+} {\rm e}^{2 iw B} \alpha(w') v(w')  \frac{dw'}{2\pi i}.
\end{equation}
Here the integration contour $C_+$ encircles 
the positive imaginary axis of $w'$ but it excludes the
pole of $v(w')$ at $w'= i 0^+$.

Set
\[
u(w)=-\frac{1+iw}{a \Kp(0)} v(w).
\]
By using the condition  (\ref{epsilonBeq0}),    one rewrites the integral equation (\ref{integeqv}) in terms of 
$u(w)$ as
\begin{align}
u(w) &=\frac{i}{w}  
+ \int_{C_+}  \frac{ {\rm e}^{2iw'B}}{w+w'+i0^+} \rho(w') u(w')  \frac{dw'}{2\pi i},  \label{eq:integraleq_u}\\
\rho(w)&=\frac{1-iw}{1+iw} \alpha(w). \nonumber
\end{align}
Here we understand that the pole at $w'=-w$ in the integrand is outside of ${C}_+$.

Then one derives from eq. (\ref{eq:integraleq_u}),
\begin{align}
\label{eq:wn_algebraic_0}
w_n &=\frac{1}{n} -\sum_{\ell=1}^{\infty} \frac{q^{\ell}}{\ell+n} b_{\ell} w_{\ell},   \qquad n \in \mathbb{N} ,
 \nn  \\
&w_n= \frac{4}{3} u(\frac{4}{3}n i),
 \\
&b_n=  \frac{(-1)^n}{n! (n-1)!} \frac{\Gamma (\frac{n}{3}) \Gamma(\frac{3}{2}+\frac{2n}{3})}{ \Gamma (-\frac{n}{3}) \Gamma(\frac{3}{2}-\frac{2n}{3})} ,
\nonumber
\end{align}
where 
\be
\label{defq}
q={\rm e}^{-8(B+\triangle)/3} \period
\ee
Although the solution to eq. (\ref{eq:wn_algebraic_0})
 is given by an infinite power series in $q$, 
it converges quickly for a large $a/m$.

The quantity of our interest is the pseudo energy, which is given by
\begin{equation}\label{EpsAnsw}
\widehat{\epsilon}_B(w) ={\rm e}^{iw B} \Kp(w) v(w) + {\rm e}^{-iw B} \Km(w) v(-w) ,
\end{equation}
and its (inverse) Fourier transformation $\epsilon_B(\theta)$.
 In the inverse Fourier integral over $\theta$ with the region  $-B<\theta<B$,  we  
 close the integration contour by adding the large semicircle in the upper half plane for the 
 first term and   in the lower half plane for the second.   
 One immediately sees that the pole of $v(w)$ at $w=i$ is canceled by 
 the zero of $\Kp(w)$.
The first term is thus evaluated by the contributions from the poles at 
$w=i 0^+, w= \frac{4}{3} n i \, (n>0, n \ne  0 \, ({\rm mod} 3) )$.
 Similarly  the second term is evaluated by those at 
 $w=i0^-, w= -\frac{4}{3} n i \, (n>0, n \ne  0 \, ({\rm mod} 3) )$.
After simple manipulations,  we obtain 
\begin{equation} \label{approx1eps}
 \epsilon(\theta) =-\epsilon_B(\theta) =-\frac{2a}{3} -\sum_{\substack{n \ge 1 \\ n\ne 0\,({\rm mod} 3)}}  c_n \cosh \frac{4 n }{3} \theta, 
\end{equation}
for $|\theta|<B$, where we have used $\Kp(0)=\sqrt{\frac{2}{3}}$.
The coefficients  $c_n$ are expressed by quantities already introduced in the above as 
\begin{align} 
c_n &=2 i {\rm e}^{-\frac{4}{3}n B}\,v_n  \,{\rm res}_{w=\frac{4}{3}n i}  \Kp , 
\nn
\\
&{\rm res}_{w=\frac{4}{3}n i}  \Kp =(-1)^{n}  \frac{4 i}{3}\sqrt{\frac{2\pi}{3}}
\frac{{\rm e}^{-\frac{4n}{3} \Delta}}{\Gamma(n) \Gamma(1-\frac{n}{3})
 \Gamma(\frac{1}{2}-\frac{2n}{3})}.   \nonumber
\end{align}
The other TBA function, $\tilde{\epsilon}=L^{-1} \tilde{\varepsilon}$, 
 is also evaluated within  the present linear approximation, which is given by
\begin{align*}
\tilde{\epsilon}(\theta) &= \sqrt{2} m \cosh \theta -\int_{-\infty}^{\infty} 
 K_2(\theta-\theta')  \epsilon(\theta') \frac{d\theta'}{2\pi}
=  \sqrt{2} m \cosh \theta +\int_{-\infty}^{\infty} \frac{\widehat{K}_2(w)}{2\pi}  
\widehat{\epsilon}_B(w) {\rm e}^{-iw\theta} \frac{dw}{2\pi}  \\
&= \sqrt{2} m \cosh \theta +\int_{-\infty}^{\infty}
 \frac{\widehat{K}_2(w) \Kp(w)}{2\pi}  v(w) ({\rm e}^{iw(B-\theta)} + {\rm e}^{iw(B+\theta)}  ) 
 \frac{dw}{2\pi}, \\
&\frac{\widehat{K}_2(w) }{2\pi} =\frac{\cosh \frac{\pi}{4}w}{\cosh \frac{\pi}{2}w}.
\end{align*}
Now we  deform the integration contour by adding a semi-circle in the upper half plane.
One easily verifies that  $\widehat{K}_2(w) \Kp(w)$ is analytic at $w=(2n+1) i$. 
It is also easy to see that the pole contribution at $ w=i $ (of $v(w)$) cancels 
 $\sqrt{2} m \cosh \theta$.
Thus, by calculating other pole contributions
at $w=i0^+$ (of  $v(w)$) and at $w=4/3n i \,(n\ne 0\, {\rm mod}\,3)$ (of $\Kp(w)$), we have
\begin{equation}\label{approx1epstilde}
\tilde{\epsilon}(\theta) =\frac{2a}{3} 
+\sum_{\substack{n \ge 1 \\ n\ne 0\,({\rm mod} 3)}} (-1)^n  c_n \cosh \frac{4 n }{3}\theta .
\end{equation}

Finally we need to represent $B$ in terms of $a$ and $m$. This is achieved by taking 
account of the pole contributions in the right hand side of  (\ref{epsilonBeq0}). 
The result is 
\begin{equation}\label{fixingB}
1- \sum_{n=1}^{\infty} \frac{b_n}{\frac{4}{3}n+1} q^n w_n = \bigl(\frac{y}{q} \bigr)^{\frac{3}{8}},
\qquad
y=\Bigl( \frac{3\sqrt{\pi} m}{2a}  \frac{\Gamma(\frac{3}{4})}{\Gamma(\frac{1}{4})} \Bigr)^{\frac{8}{3}}.
\end{equation}
This equation fixes $q$ (and thus $B$) in terms of $y$.   
It also implies  $B=\calO(L^0)$.

In appendix \ref{app:tep}, we evaluate  $\tep$ 
by directly plugging the expansion of $\ep$ (\ref{approx1eps})  into (\ref{eq:tba2dd}). 
From this, one can see explicitly how the fractional powers of $e^{\theta}$ 
appear from the summation over its integral powers,
in accordance with the periodicity required from the Y-system.
%
%+++++++++++++++++++++++++++++++++++++++++++++++++++++++++++++++++++++++++
%
 \subsection{$\calO(L^{-2})$ corrections}
 \label{sec:LTBA2}
One can extend the analysis by incorporating the $\calO(L^{-2k})$ corrections to $\epsilon$.
Here, we consider the $\calO(L^{-2})$ correction explicitly.
The results  at the leading order in this subsection reduce to those in the previous subsection.

By substituting eq. (\ref{eq:sommerfeld}) into the right hand side of  (\ref{eq:convolution3}) 
and by neglecting $\Delta_2(\theta)$,
one obtains an equation of the form  (\ref{linearALz}).
Then $Y(\theta)$ is modified by a term,
 \begin{equation}
Y(\theta)= 
\begin{cases}
X(\theta)&   \theta>B   \\
\frac{a- m{\rm e}^{\theta}}{2}+\delta_1(\theta) &  \theta<B,
\end{cases}
\nn
\end{equation}
where 
\[
 \delta_1(\theta)= -   \frac{\xi}{\cosh(\theta-B)}, \qquad
\xi= \frac{\pi}{12 L^2 |\epsilon'(B)|}.
\]
We argue in appendix \ref{app:gamma1} that $\epsilon'(B)$ is of $\calO(L^0)$. 
Thus $\delta_1$ represents the correction of $\calO(L^{-2})$ to the linearized TBA equation.
We quantitatively estimate the consequence of this modification.

Note that  
\[
\widehat{\delta}_1(w) = - \pi \xi \frac{{\rm e}^{iw B} }{\cosh\frac{\pi}{2} w}
= 2i \xi {\rm e}^{iw B} \sum_{n=0}^{\infty}(-1)^n \Bigl(
\frac{1}{w-(2n+1)i} - \frac{1}{w+(2n+1)i}
\Bigr).
\]
By using this, one can show that  $v(w)$ satisfies  eq. (\ref{integeqv}) with
\begin{align}
D_v(w)&= -i\frac{a \Kp(0)}{2(w-i 0^+)} + i \frac{m e^{B} \Kp(-i)}{2(w-i)}+ \delta_-(w){\rm e}^{-iw B}, 
\nn
\\
\delta_-(w)&=  2i \xi{\rm e}^{iw B}
 \sum_{n=0}^{\infty} (-1)^n\frac{ \Km((2n+1)i)}{w-(2n+1)i}. \nonumber 
\end{align}  
The last term is the only modification due to the leading term from $\Delta_1(\theta)$.
It however changes the analytic property of  $v(w)$. As shown previously, 
 $v(w) $ is almost pole free in the upper half plane of $w$  without  $\delta_-(w)$.
This time it  has infinitely many poles in {\it both} half planes.
  More explicitly,
\begin{equation}
\label{vw}
v(w)=  -i\frac{a \Kp(0)}{2}(\frac{1}{w-i0^+}+\frac{1}{w+i0^+})   + i \frac{m e^{B} \Kp(-i)}{2(w-i)}
+ \sum_{n\ge 1} \frac{ {\rm e}^{-\frac{8}{3} n B}}{w+\frac{4}{3}n i} \alpha_n  v(\frac{4}{3}n i)
+  \delta_-(w){\rm e}^{-iw B}. 
\end{equation}
The convolution integral in 
(\ref{integeqv}) does not produce further terms as 
$\alpha(w) \delta_-(w)$ is regular at $w=(2n+1)i,\, n \in \mathbb{Z}_{\ge 0}$.

The analogue of (\ref{epsilonBeq0})  reads,
\begin{align}\label{condtion1}
-1&+\frac{m {\rm e}^{B} \Km(i)}{2 a \Kp(0)} +\frac{\xi_1}{L^2}\zeta_c 
=
\int_{C_+} \frac{ i{\rm e}^{2iw'B}}{a\Kp(0)} \alpha(w') v(w')  \frac{dw'}{2\pi i}, \\
&\xi_1= \frac{2\xi L^2}{a \Kp(0)}, \qquad 
\zeta_c= \frac{1}{2} + \sum_{\ell=0}^{\infty} (-1)^\ell  \bigl(\Km((2\ell+1)i)-1)  .
\nonumber 
\end{align}
Here $\xi_1 =\calO(L^0)$,  and it is estimated in appendix \ref{app:gamma1}.
By utilizing this, one derives the equation for $u(w)$,
\begin{align}\label{eq:integraleq_uII}
u(w) &=\frac{i}{w} + \frac{\xi_1}{L^2} \sum_{n=1}^{\infty}   \frac{2n (-1)^n i}{w-(2n+1)i} \Km((2n+1)i)
+ \int_{C_+}  \frac{ {\rm e}^{2iw'B}}{w+w'+i0^+} \rho(w') u(w')  \frac{dw'}{2\pi i}. 
\end{align}
As mentioned previously, the pole at $w'=-w$ in the integrand is outside of  $C_+$.
From this, we again obtain a set of algebraic equations for $w_n$,
\begin{equation}\label{eq:algebraic_wn}
w_n =\frac{1}{n} + \frac{4\xi_1}{3L^2} \sum_{\ell=1}^{\infty} (-1)^{\ell} \frac{2\ell }{\frac{4n}{3}-(2\ell+1)}  \Km((2\ell+1)i)
-\sum_{\ell\ge 1} \frac{q^{\ell}}{n+\ell} b_{\ell} w_{\ell},
\end{equation}
where $q$ is defined in (\ref{defq}).
For a given sequence $\omega_{n,0} \, (n >0)$ of $\calO(1)$, let $\omega_n$ be a solution to
\be
\label{eqomega}
\omega_n=\omega_{n,0} -\sum_{\ell\ge 1} \frac{q^{\ell}}{n+\ell}b_{\ell}  \omega_{\ell}.
\ee
It has  a power series solution of the form,
\[
\omega_n=\sum_{\ell=0}^{\infty} \omega_{n,\ell} q^{\ell}.
\]
The coefficients  $\omega_{n,\ell}\, (\ell\ge 1)$ are uniquely determined by  $\omega_{n,0} $. 
The first few coefficients read explicitly
\begin{align*}
\omega_{n, 1} &= -\frac{1}{n+1} b_1 \omega_{1,0}, \qquad
\omega_{n, 2} = \frac{1}{2n+2}  b_1^2 \omega_{1,0} - \frac{1}{n+2}  b_2 \omega_{2,0}  , \\
\omega_{n, 3} &= -\frac{1}{4(n+1)} b_1^3 \omega_{1,0} 
   + \frac{1}{3(n+2)} b_1 b_2 \omega_{1,0}
   + \frac{1}{3(n+1)} b_1 b_2 \omega_{2,0} 
   -  \frac{1}{n+3} b_3 \omega_{3,0} \period
\end{align*}

The solution to (\ref{eq:algebraic_wn}) is thus written in the form,
\begin{equation}\label{eq:w_n_sum_w0_w1}
w_n = w^{(0)}_n(q) + \frac{ w^{(1)}_n(q)}{L^2}, \qquad
 w^{(0)}_n(q) =\sum_{\ell=0}^{\infty} \gamma_{n\ell} q^{\ell}, \quad  w^{(1)}_n(q) =\sum_{\ell=0}^{\infty} \kappa_{n\ell} q^{\ell}.
\end{equation} 
In the above we have used
\be
\label{cdnell}
\gamma_{n\ell}= \omega_{n,\ell}|_{\omega_{n,0} \rightarrow 1/n} \comma \qquad
\kappa_{n\ell}= \omega_{n,\ell}|_{\omega_{n,0} \rightarrow \kappa_{n0}} \comma
\ee
and
\be
\label{kappan0}
 \kappa_{n0}= \frac{4\xi_1}{3} \sum_{\ell=1}^{\infty} (-1)^{\ell} \frac{2\ell }{\frac{4n}{3}-(2\ell+1)}  \Km((2\ell+1)i).
\ee
Note that   $w^{(0)}_n(q)$  and $w^{(1)}_n(q)$  are {\it not} decompositions of  $w_n$ 
into $\calO(L^0)$ and $\calO(L^{-2})$ terms.
As will be shown below, $q$ is given by a power series in  $L^{-2}$.

Next we represent $q$ in terms of $y$ defined  in (\ref{fixingB}).
This is achieved by rewriting  (\ref{condtion1}) in terms of $w_n$,
\begin{equation} \label{condtion2}
1-\sum_{\ell= 1}^{\infty}  \frac{b_\ell}{\frac{4}{3}\ell+1} q^{\ell} w_{\ell} = (1+\frac{\nu}{L^2} ) \bigl(\frac{y}{q}\bigr)^{\frac{3}{8}}.
\end{equation}
Here we introduced
\[
\nu=\frac{2 \xi_1{\rm e}^{-B} \Kp(0) a}{\Km(i) m} \zeta_c \comma
\]
which is again of $\calO(L^0)$.

In order to solve this we set
\begin{equation}\label{eq:q_sum}
q=q^{(0)} +q^{(1)} L^{-2}+\calO(L^{-4}).
\end{equation}
By substituting  (\ref{eq:w_n_sum_w0_w1}) and (\ref{eq:q_sum}) into (\ref{condtion2})
one obtains equations of $\calO(L^0), \calO(L^{-2})$ as 
\begin{align}\label{eq:condition3}
&1-\sum_{{\ell}=1}^{\infty} \frac{b_{\ell} (q^{(0)})^{\ell} \widetilde{w}^{(0)}_{\ell}}{\frac{4}{3}{\ell}+1} 
= \bigl(\frac{y}{q^{(0)}} \bigr)^{\frac{3}{8}},
\nonumber \\
&-q^{(1)}\sum_{{\ell}=1}^{\infty}  \frac{{\ell}b_{\ell} (q^{(0)})^{{\ell}-1} 
\widetilde{w}^{(0)}_{\ell}}{\frac{4}{3}{\ell}+1
}- \sum_{{\ell}=1}^{\infty}  \frac{b_{\ell} (q^{(0)})^{{\ell}} \widetilde{w}^{(1)}_{\ell}}{\frac{4}{3}{\ell}+1}
=  \bigl(\frac{y}{q^{(0)}} \bigr)^{\frac{3}{8}} \bigl(\nu-\frac{3 q^{(1)}}{8 q^{(0)}} \bigr), 
\end{align}
respectively.   
The coefficients $\widetilde{w}^{(0)}_{\ell} $ and $ \widetilde{w}^{(1)}_{\ell}$ are given by
\be
\label{tildewq}
\widetilde{w}^{(0)}_{n}=\sum_{{\ell}=0}^{\infty} \gamma_{n{\ell}} (q^{(0)})^{\ell},
\qquad
\widetilde{w}^{(1)}_{n}=q^{(1)} \sum_{{\ell}=0}^{\infty} \ell \gamma_{n{\ell}} (q^{(0)})^{{\ell}-1}+\sum_{{\ell}=0}^{\infty} \kappa_{n{\ell}}  (q^{(0)})^{\ell}.
\ee
Note that they are now  decompositions of $w_{\ell}$ into $\calO(L^0)$ and  $\calO(L^{-2})$ 
terms,  
\[
w_{\ell}= \widetilde{w}^{(0)}_{\ell} + L^{-2} \widetilde{w}^{(1)}_{\ell} +\calO(L^{-4}).
\] 
We assume expansions
\be
\label{q01}
q^{(0)}= y+\sum_{{\ell}=2}^{\infty} q^{(0)}_{\ell} y^{\ell},\quad
q^{(1)}= \frac{8}{3} \nu y + \sum_{{\ell}=2}^{\infty} q_{\ell}^{(1)} y^{\ell} ,
\ee 
and substitute them into  (\ref{eq:condition3}). Then the coefficients
 $q^{(0)}_{\ell}$ and $q^{(1)}_{\ell}$ are determined order by order in $y$.
 For later use, we list $q^{(0)}_\ell$ for $\ell=2,3,4$:
\be
  q^{(0)}_2 =  \frac{8}{7} b_1 
  \comma \quad
  q^{(0)}_3 =  \frac{80}{49} b_1^2 +  \frac{4}{11} b_2
  \comma \quad
  q^{(0)}_4 =  \frac{382}{147} b_1^3 +  \frac{320}{231} b_1 b_2  +  \frac{8}{45} b_3 \period
  \nn
\ee
 
 The key quantity $\widehat{\epsilon}_B(w)$ 
has formally  the same form as  (\ref{EpsAnsw}).
We take its  inverse Fourier transformation.  
The extra poles of  $v(w)$ at $w=(2n+1)i\, (n \ge 0)$ do not play a role as  the combination
$ \Kp(w) v(w)$ is regular at these points. The integration contour is closed  
in the upper (lower) half plane for the first (second) term.  
We then  arrive at
\be
\epsilon(\theta)=- \frac{2a}{3}-  \sum_{\substack{n \ge 1 \\ n\ne 0\,({\rm mod} 3)}} c_n \cosh\frac{4n}{3}\theta, \qquad
c_n= 2i {\rm e}^{-\frac{4}{3}n B} v_n {\rm res}_{w=\frac{4}{3}ni} \Kp(w).
\nn
\ee
This is formally identical to the one without $\Delta_1(\theta)$.
The coefficients  $v_n $, however, contain  higher order terms of $\calO(L^{-2k})$
and the coefficients $c_n$ also contain them similarly.
The $\calO(L^{0})$ and $\calO(L^{-2})$  terms are 
given explicitly in terms of other coefficients obtained
so far,
\begin{align}
\label{cn01}
c_n &= \bigl(c_n^{(0)}+ \frac{ c_n^{(1)}}{L^2} \bigr)  a +\calO\Bigl(\frac{1}{L^4}\Bigr),  \nn \\
&c^{(0)}_n = (q^{(0)})^{\frac{n}{2}} \widetilde{w}^{(0)}_n  
\widetilde{c}_n  \frac{3 \Kp(0) }{4 (\frac{4}{3}n-1)}, \qquad 
c^{(1)}_n = (q^{(0)})^{\frac{n}{2}} \bigl( \widetilde{w}^{(1)}_n +\frac{n q^{(1)}}{2 q^{(0)}} 
 \widetilde{w}^{(0)}_n \bigr)  \widetilde{c}_n  \frac{3 \Kp(0) }{4 (\frac{4}{3}n-1)}, \nn\\
&\widetilde{c}_n   = \frac{8}{3} \sqrt{\frac{2\pi}{3}} \frac{(-1)^{n+1}}
{\Gamma(n) \Gamma(1-\frac{n}{3}) \Gamma(\frac{1}{2}-\frac{2n}{3})}.
\end{align}
By using these, we have
\be
\label{epexp}
\frac{\epsilon(\theta)}{a} = -\frac{2}{3}-  \sum_{\substack{n \ge 1 \\ n\ne 0\,({\rm mod} 3)}} 
c^{(0)}_n \cosh\frac{4n \theta}{3}
- \frac{1}{L^2}  \sum_{\substack{n \ge 1 \\ n\ne 0\,({\rm mod} 3)}} c^{(1)}_n \cosh\frac{4n\theta}{3} 
+\calO\Bigl( \frac{1}{L^{4}}\Bigr).
\ee

We can similarly estimate the correction to $\tilde{\epsilon}(\theta)$,
\begin{equation}\label{eq:varepsL2}
\tilde{\epsilon}(\theta) = \sqrt{2} m \cosh \theta -\int_{-\infty}^{\infty} K_2(\theta-\theta') 
\epsilon(\theta') \frac{d\theta'}{2\pi}
+\xi (K_2(\theta-B)+ K_2(\theta+B)).
\end{equation}
The convolution term is again given by 
\[
\int_{-\infty}^{\infty} \frac{\widehat{K}_2(w) \Kp(w)}{2\pi}  v(w) ({\rm e}^{iw(B-\theta)} 
+ {\rm e}^{iw(B+\theta)}  ) \frac{dw}{2\pi}. \\
\]
As $|\theta|<B$ we close the contour in the upper half plane.
There are contributions from the poles 
at  $w=i0^+$, $w=i$ and $w=\frac{4}{3}n i \, (n\ne 0 \, ({\rm mod}\, 3))$ as before.
In addition, there appear additional contributions at   $w=(2n+1) i$, 
originated from the last term in  (\ref{vw}).
After a simple calculation we find that the additional  term cancels the last two terms in 
(\ref{eq:varepsL2}),
\[
\xi (K_2(\theta-B)+ K_2(\theta+B))-\sum_{n=0} 8 \xi (-1)^n 
\sin \frac{2n+3}{4} {\rm e}^{-(2n+1) B}  \cosh (2n+1) \theta
=0.
\]
Thus we have
\begin{align}
\label{tepexp}
\frac{\tilde{\epsilon}(\theta)}{a} 
=\frac{2}{3} 
+ \sum_{\substack{n \ge 1 \\ n\ne 0\,({\rm mod} 3)}} (-1)^n c^{(0)}_n \cosh\frac{4n \theta}{3}
+ \frac{1}{L^2}  
\sum_{\substack{n \ge 1 \\ n\ne 0\,({\rm mod} 3)}}(-1)^n  c^{(1)}_n \cosh\frac{4n\theta}{3}
 +\calO\Bigl( \frac{1}{L^{4}}\Bigr). 
\end{align}
This and (\ref{epexp}) are
 consistent with the periodicity property of the Y-system (\ref{period6}).
%
%--------------------------------------------------------------------------------
%
\subsection{Free energy}
\label{sec:LTBAf}

The free energy $A_{\rm free}$  is  represented in terms of the TBA functions as
in (\ref{Afree}).
As before, we drop the terms which are exponentially small 
by assuming the scaling  (\ref{eq:epscale}), and write
\be
\label{Afree2}
A_{\rm free}=- \Delta {\cal{E}} +\Delta_F,
\ee
where
\begin{align*}
\Delta {\cal{E}} &= \int^B_{-B} M \cosh \theta \varepsilon(\theta) \frac{d\theta}{2\pi}=
- \int^{\infty}_{-\infty} m L^2 \cosh \theta \epsilon_B (\theta) \frac{d\theta}{2\pi},   \\
\Delta_F &= \bigl( \int_{|B-\theta|<\eta} +  \int_{|B+\theta|<\eta} \bigr)
m L \cosh \theta  \log(1+{\rm e}^{-|\varepsilon(\theta)|})  \frac{d\theta}{2\pi}.
\end{align*}

For the moment, we concentrate on  $\Delta {\cal{E}}$.
One can show that the following relation, derived in \cite{AlZamolodchikov1995},
 is still valid even in the presence of $\Delta_1(\theta)$, 
\[
\Delta {\cal{E}} =-\frac{m L^2}{2\pi} \widehat{\epsilon}_B(-i).
\] 
Thanks to the integral equation for  $\widehat{\epsilon}_B$, we have 
\[
\Delta {\cal{E}} = -\frac{m L^2 {\rm e}^{B} \Km(i)}{2\pi} \Bigl(
D_v(-i) +\int_{C'_+} \frac{{\rm e}^{2iwB}}{w-i} \alpha(w) v(w) \frac{dw}{2\pi i } \Bigr).
\]
Here the contour $C'_+$ encircles the positive imaginary axis, which includes $w=i$.
We use the  explicit form of $D_v(w)$ and the ``boundary condition" (\ref{condtion1}) 
to represent this as
\[
\Delta {\cal{E}} = -\frac{m a L^2 {\rm e}^{B} \Km(i) \Kp(0)}{4\pi}
\Bigl( 1- \int_{C_+} \frac{dw}{2\pi i} {\rm e}^{2iw B} \frac{\rho(w) u(w)}{w-i}
+ \frac{\xi_1}{L^2} \sum_{n=0}^{\infty} (-1)^n \frac{n}{n+1} \Km((2n+1)i) \Bigr).
\]

We  write  $\Delta {\cal{E}}=\Delta {\cal{E}}_1+\Delta {\cal{E}}_2$
where  $\Delta {\cal{E}}_1$ stands for the  contribution from the pole
at $w=i$ in the integral. Explicitly,
\[
\frac{\Delta {\cal{E}}_1}{L^2} =\frac{m^2}{4} +\frac{am}{2}{\rm e}^{-B} \Kp(0) \frac{\xi_1}{L^2}. 
\]
Here we used 
\[
u(i) =\frac{-i}{a\Kp(0)} {\rm res}_{w=i}v(w) =
 \frac{\Km(i)}{a\Kp(0)}( \frac{m{\rm e}^{B}}{2} +2  \xi ).
\]

Similarly, 
\begin{align*}
\frac{\Delta {\cal{E}} _2}{L^2} =& -\frac{ (a \Kp(0))^2}{2\pi}
 \Bigl(   u(i) -\frac{\xi_1}{L^2}  \Km(i) \Bigr )  \\
&\times  \Bigl(  1- \int_{C'_+\backslash\{i\}} \frac{{\rm e}^{2iwB}}{w-i} \rho(w) u(w) \frac{dw}{2\pi i }
+  \frac{\xi_1}{L^2} \sum_{n=0}^{\infty}(-1)^n \frac{n}{n+1} \Km( (2n+1)i)     \Bigr).
\end{align*}
The content of the first bracket in the right hand side is evaluated by use of 
(\ref{eq:integraleq_uII}),
\begin{align}
\label{alpha01}
&u(i) -\frac{\xi_1}{L^2} \Km(i) = 1-\sum_{n=1} \frac{3}{4n+3} q^n b_n w_n
 - \frac{\xi_1}{L^2} \zeta_c   =\alpha^{(0)} + \frac{\alpha^{(1)}}{L^2} 
 + \calO\Bigl(\frac{1}{L^4}\Bigr), \nn \\
&\alpha^{(0)}=1-\sum_{n=1}^{\infty} \frac{3}{4n+3} (q^{(0)})^n b_n \widetilde{w}^{(0)}_n, \\
&\alpha^{(1)}=-\sum_{n=1}^{\infty}  \Bigl(\frac{3}{4n+3} (q^{(0)})^n b_n \widetilde{w}^{(1)}_n
+   \frac{3n}{4n+3} (q^{(0)})^{n-1}  q^{(1)} b_n \widetilde{w}^{(0)}_n  \Bigr)
-\xi_1 \zeta_c. \nn 
\end{align}

The second bracket term is also decomposed into  
$\beta^{(0)} +\beta^{(1)} L^{-2} + \calO(L^{-4})$ where
\begin{align}
\label{beta01}
&\beta^{(0)}=1+\sum_{n=1} \frac{3}{4n-3} (q^{(0)})^n b_n \widetilde{w}^{(0)}_n , \nn  \\
&\beta^{(1)}=\sum_{n=1} \Bigl(\frac{3}{4n-3} (q^{(0)})^n b_n \widetilde{w}^{(1)}_n
+   \frac{3n}{4n-3} (q^{(0)})^{n-1}  q^{(1)} b_n \widetilde{w}^{(0)}_n  \Bigr)
+\xi_1 (\zeta_c -\zeta'_c) , 
\end{align}
with 
\be
  \zeta'_c = \log 2 + \sum_{\ell=0}^\infty \frac{(-1)^\ell}{\ell+1} \bigl( \Km((2\ell+1)i)-1 \bigr) \period
  \nn
\ee
Thus we have
\begin{align}
\label{DelEp}
\frac{\Delta {\cal{E}}}{L^2}  = & \Delta {\cal E}^{(0)} + \frac{\Delta {\cal E}^{(1)}}{L^2} 
    + \calO\Bigl(\frac{1}{L^{4}}\Bigr) \comma \nn \\
&  \Delta {\cal E}^{(0)} = \frac{m^2}{4}  -\frac{ (a\Kp(0) )^2}{2\pi} \alpha^{(0)} \beta^{(0)}
 \comma \\
 &  \Delta {\cal E}^{(1)} = 
   \frac{am \xi_1}{2}{\rm e}^{-B} \Kp(0) 
  -\frac{ (a\Kp(0) )^2}{2\pi}  \bigl(\alpha^{(0)} \beta^{(1)}+\alpha^{(1)} \beta^{(0)} \bigr) 
  \period \nn
\end{align}

 As for $\Delta_F$, the lowest order approximation gives
\begin{align}
\Delta_F &=  \frac{\pi}{12 |\epsilon'(B)|}  2 m \cosh B + \calO \Bigl(\frac{1}{L^{2}}\Bigr)
=  \delta_F + \calO\Bigl(\frac{1}{L^{2}}\Bigr)  \comma 
\nn
\end{align}
where 
\be
\label{deltaF}
 \delta_F = \frac{\pi m}{12 |\epsilon'(B)|}  
 \Bigl( (q^{(0)})^{\frac{3}{8}} e^{\Delta} +  (q^{(0)})^{-\frac{3}{8}} e^{-\Delta} \Bigr)  \period
\ee
We have substituted  (\ref{defq})  and (\ref{eq:q_sum}) into $e^B$.
Thus, we obtain
\be
\label{Afree3}
  A_{\rm free} =  -L^2 \Delta {\cal E}^{(0)} - \Delta {\cal E}^{(1)} + \delta_F
 +   \calO\Bigl(\frac{1}{L^{2}}\Bigr) \period
\ee

%%%
%
\subsection{Leading order expansions in $1/L$}
\label{sec:MAexp}

In section \ref{sec:RExpansion}, we will apply the results in this section 
for the 6-point MHV amplitudes at strong coupling.
To express  the expansions of $\ep, \tep$ and $A_{\rm free}$,
and hence the amplitudes in terms of the
parameters in the TBA equations $M, A$ or $y$ in (\ref{fixingB}), 
we first solve (\ref{eqomega}) to find $ \tilde{w}_n^{(0)}, \tilde{w}_n^{(1)}$
 in (\ref{tildewq}). We then solve (\ref{eq:condition3}) 
to find $ q^{(0)}, q^{(1)}$ in terms of $y$, and substitute them 
into (\ref{epexp}), (\ref{tepexp}) and (\ref{Afree3}) with (\ref{cn01}), (\ref{alpha01}), 
(\ref{beta01}), (\ref{DelEp}) and (\ref{deltaF}). 

Since we are interested in the expansions for large $L$, 
we will focus in section \ref{sec:RExpansion} on the expansions 
at the leading order in $1/L^2$.
In this case, one needs $\tilde{w}_n^{(0)}$ and $q^{(0)}$ only
and does not need to take into account $c_n^{(1)}, \Delta {\cal E}^{(1)}$ and $\delta_F$.
We then have  simple expansions of $\ep, \tep$:
\eqb
\label{epexp2}
 \ep(\theta) \Eqn{=}
     a \sqrt{\pi} \cdot  \sum_{n=0}^\infty e_n(\theta)  y^{n/2} +\calO(L^{-2}) \comma \nn \\
 \tep(\theta) \Eqn{=} a \sqrt{\pi} \cdot  \sum_{n=0}^\infty (-1)^{n+1} e_n(\theta) y^{n/2} 
  +\calO(L^{-2}) \comma
\eqe
for $ |\theta|  < B $,
where the coefficients $e_n$ for small $n$ are found to be
\begin{align}
   & e_0(\theta) {=} -\frac{2}{3\sqrt{\pi}} \comma &
   & e_1(\theta) {=} -\frac{4 \cosh\frac{4}{3} \theta}{\Gamma(-\frac{1}{6})\Gamma(\frac{2}{3})} \comma
   \nn \\
   & e_2 (\theta){=}  \frac{\frac{2}{5} \cosh\frac{8}{3}\theta}{\Gamma(-\frac{5}{6})\Gamma(\frac{1}{3})} 
    \comma &
    & e_3(\theta) {=}  -\frac{\frac{2}{7}b_1 \cosh\frac{4}{3}\theta}{\Gamma(-\frac{1}{6})\Gamma(\frac{2}{3})}
     \comma \nn  \\
    &e_4(\theta) {=}  \frac{\frac{1}{78} \cosh\frac{16}{3}\theta}{\Gamma(-\frac{13}{6})\Gamma(-\frac{1}{3})}  
         + \frac{\frac{4}{21} b_1 \cosh\frac{8}{3}\theta}{\Gamma(-\frac{5}{6})\Gamma(\frac{1}{3})} 
         \comma  &
   & e_5(\theta) {=}   
   -\frac{\frac{1}{510} \cosh\frac{20}{3}\theta}{\Gamma(-\frac{17}{6})\Gamma(-\frac{2}{3})}  
      -  \frac{\bigl(\frac{9}{49} b_1^2 + \frac{2}{33}b_2\bigr) \cosh\frac{4}{3}\theta}{
      \Gamma(-\frac{1}{6})\Gamma(\frac{2}{3})} 
      \comma  \nn \\
     &&& e_6 (\theta) {=}  \frac{\frac{2}{105} b_1 \cosh\frac{16}{3}\theta}{
     \Gamma(-\frac{13}{6})\Gamma(-\frac{1}{3})}  
      + \frac{\bigl(\frac{26}{147} b_1^2 + \frac{1}{22}b_2\bigr) \cosh\frac{8}{3}\theta}{
      \Gamma(-\frac{5}{6})\Gamma(\frac{1}{3})}
      \comma  & \nn
\end{align}
with $b_n$  given in (\ref{eq:wn_algebraic_0}).

Similarly, the leading order term of the free energy is
%$f$ is
\cite{AlZamolodchikov1995} 
\be
\label{fexp}
   f  := -A_{\rm free}/L^2
   = \Delta {\cal E}^{(0)} +\calO(L^{-2})  \comma \qquad
    \Delta {\cal E}^{(0)} = \frac{m^2}{4}   - \frac{a^2}{\pi}  \sum_{n=0}^\infty k_n y^n 
  \comma
\ee
where we have introduced $f$ so that the leading terms are of $\calO(L^0)$.
The coefficients $k_n$ for small $n$ are%
\footnote{
We correct typos in $k_3, k_4$ in \cite{AlZamolodchikov1995}.
}
\begin{align}
\label{fk}
  & k_0 =  \frac{1}{3} \comma & & k_1 = \frac{6}{7} b_1   \comma \nn \\
  & k_2 =  \frac{6}{49} b_1^2 +  \frac{3}{55} b_2 \comma 
     & & k_3 = \frac{57}{686} b_1^3 + \frac{6}{77} b_1b_2 + \frac{2}{135} b_3   \comma \\
  &&& k_4 =    \frac{29}{343} b_1^4 + \frac{4}{33} b_1^2 b_2 + \frac{3}{242} b_2^2 
         + \frac{4}{105} b_1 b_3 + \frac{3}{494} b_4 \period \nn
\end{align}

The $\calO(y^0)$ terms agree with the UV limits in (\ref{smallM}).
Though these expansions are  based on the TBA equations of the form (\ref{TBA6}) which are
valid for $|\! \im \theta  | < \pi/4 $, they are analytically continued for any $\im \theta$. 

\par\smallskip
\ni
{\bf Overlap with the small mass expansion :}
%
%Though we have discussed the case where both $A$ 
%and $M$ are large, the results may be continued to small $M$, or those 
%of the small $M$ expansion may be continued to the regime of our expansion.
%Thus as a non-trivial check of these expansions, they can be compared 
%with the results of the small $M$ expansions with other parameters fixed 
%\cite{Hatsuda:2014vra}.
%Indeed, taking into account the relation of the notation here and that in \cite{Hatsuda:2014vra},
Though we have discussed the case where both $A$ 
and $M$ are large, the results may be continued to small $M$, or those of
the small $M$ expansions with other parameters fixed 
\cite{Hatsuda:2014vra} may be continued to the regime of our expansions.
Thus as a non-trivial check, they can be compared.
%with the results of the small $M$ expansions with other parameters fixed 
%\cite{Hatsuda:2014vra}.
Indeed, taking into account the relation of the notation here and that in \cite{Hatsuda:2014vra},
\be
 \log Y_{\rm 1, there}(\theta + i \varphi) 
  = \varep(\theta) + A \comma \quad
   \ell_{\rm there} = M \comma \quad 
   \phi_{\rm there} = \frac{2}{3i} A \comma \nn
\ee
we find that the expansions of $Y_{\rm 1, there}$ and $A_{\rm free}$ in 
\cite{Hatsuda:2014vra} precisely reproduce for large $A$ the leading terms 
of our expansions,
\be
  a \sqrt{\pi} e_1(0) y^{1/2}   \comma \quad
 -\frac{a^2}{\pi} k_1 y  \period \nn
\ee

%%%
\section{Numerical analysis}

In the previous section, the TBA equations for the 6-point amplitudes (\ref{TBA6})
for large $A$ and $M$ were shown to reduce to (\ref{eq:convolution3}) up to exponentially
small terms in $L \sim A$. 
The corrections from  $\Delta_1$ are of  $\calO(L^{-1})$ and hence relatively of $\calO(L^{-2})$,
and those from $\Delta_2$ are again exponentially small from the assumption on
the scaling of the pseudo energies (\ref{eq:epscale}). 
Once these corrections are dropped,
the TBA equations further reduce to the linearized TBA equations
(\ref{eq:tba1dd}) and (\ref{eq:tba2dd}).
These are solved by the method in \cite{AlZamolodchikov1995}.
By extending the analysis in \cite{AlZamolodchikov1995},
the relatively $\calO(L^{-2})$ terms for $\varep, \tvarep $ were 
explicitly evaluated. Similarly, the free energy $A_{\rm free}$
reduces to (\ref{Afree2}) up to 
terms which are exponentially small or expectedly exponentially 
small by  the scaling  (\ref{eq:epscale}).
The leading contributions are of 
$\calO(L^2)$, and the others are of $\calO(L^0)$ and hence relatively of $\calO(L^{-2})$  again.
By following and extending the analysis in \cite{AlZamolodchikov1995}, 
we evaluated both contributions explicitly. 

\subsection{Dependence on $L$}
Now, we confirm the $L$-dependence of the expansions  (\ref{epexp}), (\ref{tepexp})
and (\ref{Afree3}).

This also serves as a check of our assumption 
on the scaling (\ref{eq:epscale}), which assures that $\Delta_2$ and a similar term dropped in 
(\ref{Afree2}) are indeed exponentially small in $L$.
For this purpose, we first consider the pseudo energy at the special value of $\theta=0$.
From the results in the previous section, we have
\be
\label{ep0L}
 \epsilon(0) = a_1 + \frac{a_2}{L^2} + \calO(L^{-4}) \comma 
\ee
up to (expectedly) exponentially small terms with
\be
a_1 = 
   a\sqrt{\pi} \cdot  \sum_{n=0}^\infty e_n(0)  y^{n/2} \comma \qquad 
a_2 = -  a\sum_{\substack{n \ge 1 \\ n\ne 0\,({\rm mod} 3)}}  c^{(1)}_n . \nn
\ee
One has a similar  expression for $\tep(0)$.
As for the free energy, the $L$-dependence reads
\be
\label{fL}
 f = d_1 + \frac{d_2}{L^2}  + \calO(\frac{1}{L^{4}}) \comma 
\ee
up to (expectedly) exponentially small terms, where $d_1 = \Delta {\cal E}^{(0)}$ in (\ref{fexp}) 
and $d_2 =  \Delta {\cal E}^{(1)} -\delta_F$  in (\ref{Afree3}).

Table \ref{tab:parameters} summarizes the numerical values of $a_1, a_2$ and $d_1, d_2$, 
e.g. for $a=1$ and $m=10^{-\ell/4}$ $(\ell=1,...,4)$.
Here, terms up to order $y^{3} \sim (m/a)^{8}$ are included.
To evaluate $\kappa_{n0}$ in (\ref{kappan0}), 
$\Km((2\ell+1)i)$  is rewritten as $ \bigl(\Km((2\ell+1)i) -1\bigr) +1$ 
and the remainder is separated as in $\zeta_c, \zeta'_c$, so that 
the quantity in the parentheses is vanishing for large $\ell$.
$\zeta_c, \zeta'_c$ and $\kappa_{n0}$ 
are estimated by including terms up to $\ell=10^4$ in the summation.
As we have truncated the expansion, the correction is expected to be small for small $m$.

\begin{table}[H]
  \centering
  \begin{tabular}{|l||l|l||l|l|}
\hline
   $m $ &  $a_1$ & $a_2$ &  $d_1$& $d_2 $  \\
 \hline
 \hline  
 $10^{-1/4}$  &  -0.35979817 
                 & 0.34641 
                &   -0.055404898 
                &  -0.63726 
          \\
  $10^{-2/4}$ &  -0.52340676 
                     &  0.15038  
                     & -0.087142723 
                     & -0.54386 
                     \\
  $10^{-3/4} $ & -0.5999064791 
                      &  0.070976 
                      &   -0.09949568311 
                      &   -0.52780  
                      \\
  $10^{-1}$   & -0.63561331932 
                    &   0.033492 
                    & -0.103882817866 
                    &  -0.52450 
                       \\
\hline
\end{tabular}
\caption{Parameters  for the $L$-dependence of $\ep(0)$ and $f$ 
from the results in section \ref{sec:LTBA}.}
 \label{tab:parameters}
\end{table}

We compare these  with numerical results.
To this end, we have solved the original TBA equations (\ref{TBA6}) 
numerically by iteration 
as described below (\ref{kernels}). The free energy $f$ 
has also been evaluated numerically by (\ref{Afree}).
We set $a =1, m=10^{-\ell/4}$ $(\ell=1,...,4)$, 
and $L = L_k :=10^{\frac{1}{2} + \frac{k}{40}}$ 
$(k=0, ..., 100)$.
We also set the cut off of the integration over $\theta$ 
at $\theta = \pm \theta_{\rm max} $ with 
$\theta_{\rm max} = 10\cdot | \log\frac{2}{M}(A+308)|$.
The interval $|\theta | \leq \theta_{\rm max}$
is discretized by $N=10^{16}$ points.
The iteration stops
when the change of the free energy (\ref{Afree}) by one iteration becomes $2\pi \cdot 10^{-16}$,
or the number of the iterations reaches $n_{\rm max} = 100$.
 
The numerical data $\ep_{\rm num}(0)$  
of $\ep(0)$ can be  fitted by the function 
$\ep_{\rm fit}(0) = a_1 + a_2/L^{a_3}$. For the data,   e.g. from  
$L=L_k$ with $k=71, ..., 100$ and $m=10^{-3/4}$,
we obtain
\be
\label{fitep}
  a_1 = -0.5999064788  \comma \quad 
  a_2 = 0.07104736981 \comma \quad 
  a_3 =  2.000603100  \period 
\ee
A similar fitting works also for $\tep(0)$.
As for the free energy, 
the data $f_{\rm num}$ 
of $f$ for the same $L_k$ and $m$ are 
fitted by
$f_{\rm fit} = d_1 + d_2/L^{d_3}$ with 
\be
\label{fitf}
  d_1 = -0.099495683121 \comma \quad
  d_2 =  -0.5264772531 \comma \quad
  d_3 =    2.000003585 \period
\ee
These are consistent with  (\ref{ep0L}), (\ref{fL})
and the values of $(a_1, a_2)$ and $(d_1, d_2)$ in Table \ref{tab:parameters}.
 
\begin{figure}[t]
\vspace{5ex}
 \begin{center}
   \begin{minipage}{0.4\hsize}
   \begin{center}
  \includegraphics[width=73mm]{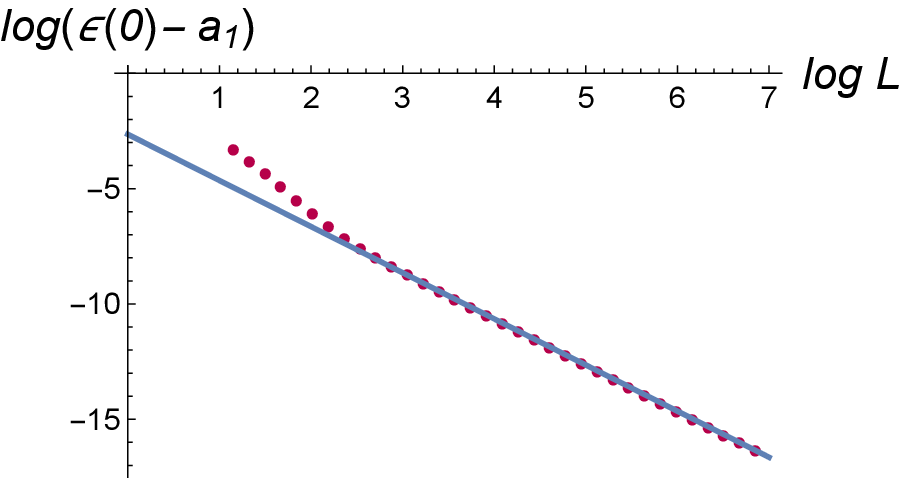}  \\ { \small (a)}
 \end{center}
 \end{minipage}
\hspace*{10ex}
\begin{minipage}{0.4\hsize}
 \begin{center}
  \includegraphics[width=71.2mm]{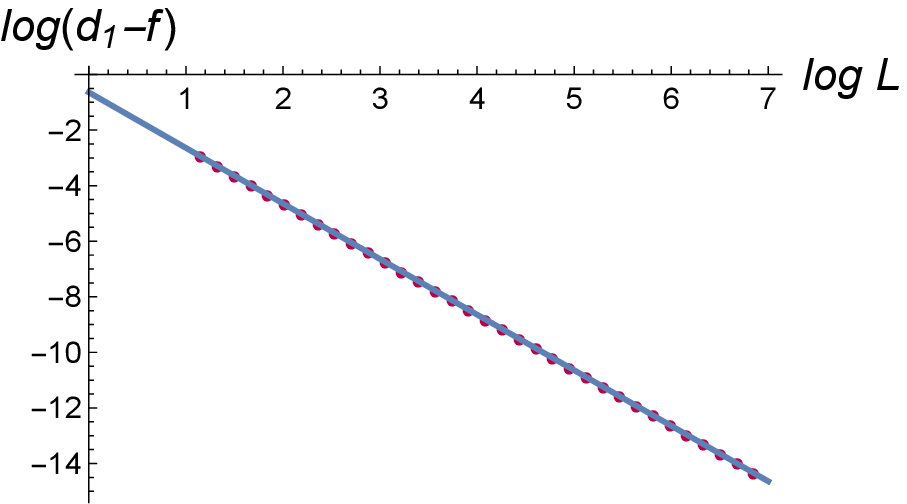}  \\ { \small (b)}
 \end{center}
 \end{minipage}
 \hfill
\caption{
(a) Plot of $\ep(0)$. (b) Plot of  $f$.
The dots represent the numerical data $\ep_{\rm num}(0), f_{\rm num}$,
whereas the solid lines represent the fitting functions $\ep_{\rm fit}(0), f_{\rm fit}$.
 }
 \label{fig:Scale}
\end{center}
\end{figure}

Figure \ref{fig:Scale} (a) and (b) are log-log plots of $\ep(0)$ and $f$, respectively.
 In (a), the dots represent the numerical data, $\log(\ep_{\rm num}(0) - a_1)$.
 The solid line represents  the fitting function,
 $\log(\ep_{\rm fit}(0) - a_1) =  \log(a_2/L^{a_3})$. The coefficients $a_i$ are given in (\ref{fitep}).
 In (b), 
 the dots represent the numerical data, $\log(d_1-f_{\rm num}(0) )$.
 The solid line represents  the fitting function,
 $\log(d_1-f_{\rm fit} ) =  \log(d_2/L^{d_3})$. The coefficients $d_i$ are given in (\ref{fitf}).
The numerical data are well fitted  for large $L$.
For small $L$, the obvious deviation of   $\ep_{\rm num}(0)$ and $\ep_{\rm fit}(0)$  
in Figure \ref{fig:Scale}  (a) is due to 
the corrections of $ \calO(e^{- ({\rm const.}) L})$.
The deviation becomes smaller for larger $m/a$, meaning that the constant in the
exponent increases as $m/a$. For the above set of the parameters, $y \sim 10^{-2}$
and $L^{-2} \geq 10^{-6}$. Since  $e_n$ and $k_n$ are also small numbers,
the truncation of the sums for $a_1$ and $d_1$
may not affect in observing the $L^{-2}$ scaling as long as the terms 
up to  $\calO(y^3)$ are included. 

By similar fittings for other $m$ with other parameters fixed, we obtain $a_j, d_j$
in Table \ref{tab:fitparameters}.
They are again consistent with the values in Table \ref{tab:parameters}.
If higher order terms in $y$ are included in evaluating these values,
we find a better agreement between Table \ref{tab:parameters}  and \ref{tab:fitparameters}.
Relative corrections for $a_2, d_2$ are larger than those
for $a_1, d_1$.
All in all, we conclude that the scaling (\ref{eq:epscale}) and 
the $L$-dependence of the expansions
are consistent with the numerical results.

\begin{table}[H]
  \begin{tabular}{|l||l|l|l||l|l|l|}
\hline
   $m $ &  $a_1$ & $a_2$ & $a_3$ &$d_1$& $d_2 $ & $d_3$ \\
 \hline
 \hline  
 $10^{-1/4}$  &  -0.35978636 
                 & 0.32640 
                 & 2.00074 
                &   -0.055407648 
                &  -0.59786 
                & 2.0001207 
          \\
  $10^{-2/4}$ &  -0.52340670 
                     &  0.14909 
                     &  2.00060  
                     &  -0.087142728 
                     & -0.53736 
                     & 2.0000179 
                     \\
  $10^{-3/4} $ &  -0.5999064788  
                      & 0.071047  
                      & 2.00060  
                      &   -0.09949568312 
                       &   -0.52648 
                      & 2.0000036 
                      \\
  $10^{-1}$   &-0.63561331929 
                    &  0.033594 
                    & 2.00061 
                    & -0.103882817867 
                    &  -0.52421 
                    & 2.0000008 
             \\
\hline
\end{tabular}
\caption{Fitting parameters for the $L$-dependence of $\ep(0)$ and $f$}
 \label{tab:fitparameters}
\end{table}

%%%%%%%%
%  figures for scaling
%%%%%%%%%
%
%%%%%

%
\subsection{Dependence on $m/a$}
\label{subsec:madep}

Next, we would like to check the expansions 
(\ref{epexp}), (\ref{tepexp}), (\ref{Afree3}) 
and (\ref{epexp2}), (\ref{fexp}).
For this purpose, we have solved  the original TBA equations (\ref{TBA6})
and evaluated the free energy (\ref{Afree}) numerically.
The parameters are set as $a =4/7, m/a=6/5 - k/100$ $(k=0, ...,115)$, $L  =10^{3}$,
and others are the same as in the previous subsection. 
Though the expansions are valid for $m/a < 1$, we include $m > a$ for reference.
It is not easy to  determine many coefficients in the expansions 
by data fitting. 
Instead, we check below if the numerical data are well approximated by 
the expansions in the previous section.

\begin{figure}[t]
\vspace{5ex}
 \begin{center}
   \begin{minipage}{0.4\hsize}
   \begin{center}
  \includegraphics[width=70mm]{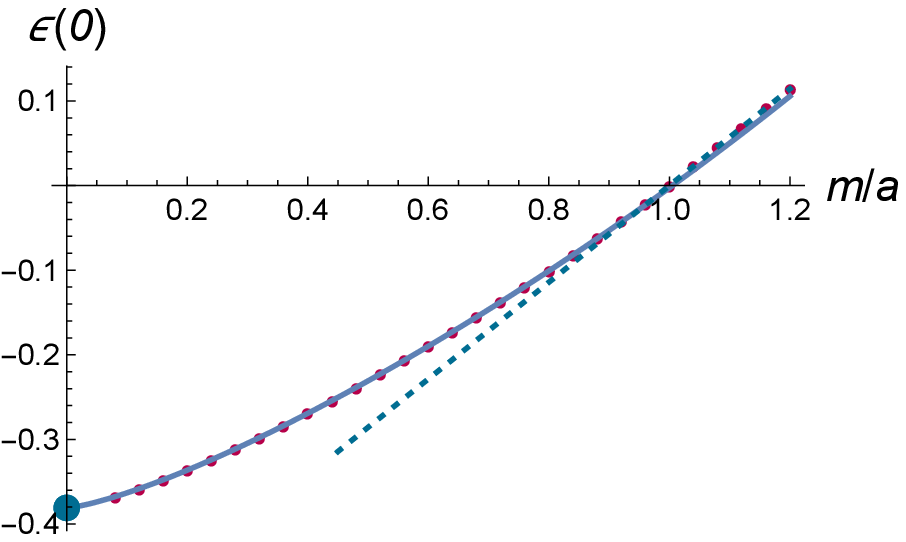}  \\ { \small (a)}
 \end{center}
 \end{minipage}
\hspace*{10ex}
\begin{minipage}{0.4\hsize}
 \begin{center}
  \includegraphics[width=70mm]{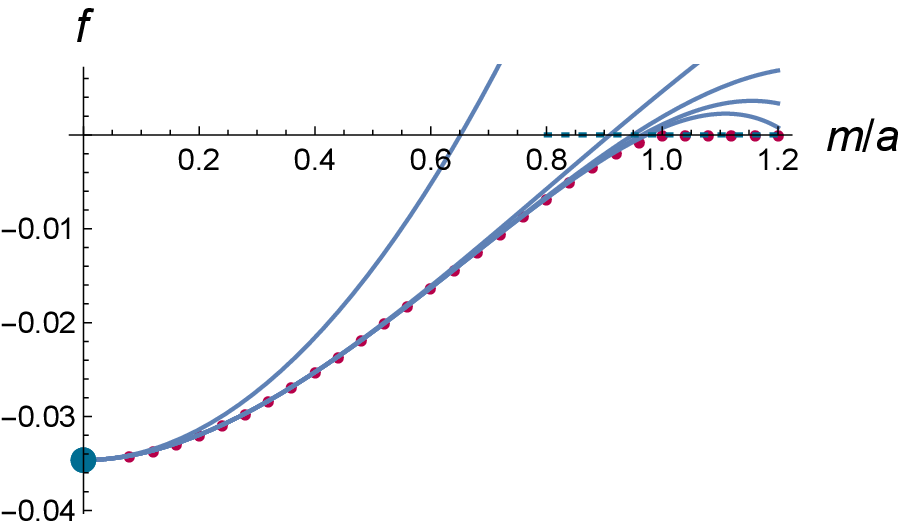}  \\ { \small (b)}
 \end{center}
 \end{minipage}
 \hfill
\caption{
 (a) Plot of $\ep(0) $. 
 The dots represent the numerical data $\ep_{\rm num}(0)$,
 whereas the solid line represents the expansion $ \ep^{(6)}(0)$.
 The blob indicates the UV value $\ep_{\rm UV}$.
  The dotted line represents  $\ep_{\rm IR}(0) $.
 (b) Plot of $f $.
 The dots represent the numerical data $\ f_{\rm num}$,
 whereas the solid lines represent the expansions  $f^{(k)}  (0 \leq k \leq 4)$ 
 from the top to the bottom.
 The blob indicates the UV value $ f_{\rm UV}$.
 The data converge to $f_{\rm IR} =0 $ (dotted line) as $m/a \to 1$.
 }
 \label{fig:madep}
\end{center}
\end{figure}

Figure \ref{fig:madep} (a) is a plot of $\ep(0)$. 
The numerical data $\ep_{\rm num}(0)$ (dots) are well approximated by $\ep^{(6)}(0)$ (solid line),
where $\ep^{(k)}(\theta)$ is the expansion (\ref{epexp2}) truncated at $n=k$. 
If other $ \ep^{(k)}$ $(k \geq 1)$ were plotted, they would be almost degenerate in the figure;
$\ep^{(1)}(0)$ is already a good approximation because $ | e_j/e_1| \ll 1$ $(1 <j)$.
As $m/a \to 0$, they approach the  value in the UV limit 
$\ep_{\rm UV} = L^{-1} \log(e^{-2A/3} + e^{-4A/3}) \approx -0.381$ (blob). 
As $m/a \to 1$, the data approach the asymptotic form 
$ \ep_{\rm IR}(0)  = -a + m$ in the IR limit (dotted line).
See (\ref{smallM}) and (\ref{largeM}).
One can also check that $ \ep_{\rm num}(0) - \ep^{(k-1)}(0)$ scales as  $ (m/a)^{4k/3}$,
and that it is saturated as $(m/a)^{4/3} L^{-2}$ for small enough $m/a$.
A similar analysis is possible also for $\tep(0)$.
including the $\calO(L^{-2})$ term.

Figure \ref{fig:madep}  (b) is a plot of $f$. 
The numerical data $f_{\rm num}$ (dots) are well approximated by $f^{(k)}$ $(k=0,1,2,3,4)$ 
(solid lines from the top to the bottom),
where $f^{(k)}$ is the expansion (\ref{fexp}) truncated at $n=k$. 
As $m/a \to 0$, they approach  the value in the UV limit
$f_{\rm UV} = -(\frac{\pi}{6} + \frac{A^2}{3\pi})/L^2 \approx -0.0346$ (blob). 
As $m/a \to 1$, the numerical data converge to the value in the IR (free field) limit
$f_{\rm IR} =0$ (dotted line).
One can also check that $ f_{\rm num} - f^{(k-1)}$ scales as  $ (m/a)^{8k/3}$,
and that it is saturated as $ L^{-2}$ without any power of $m/a$ for small enough $m/a$.
These observations are again consistent with the results in the previous section.

%%%%%%%%
%  figures for m/a dependence
%%%%%%%%%
%
%%%%%

%%%
\section{Analytic expansion of 6-point remainder function}
\label{sec:RExpansion}

Having checked the validity of the expansions and 
the linearlization against numerical data,
we now apply the expansions to the strong-coupling amplitudes.
In the following, we focus on the leading expansions in $1/A^2$,
though it is straightforward to include the subleading $\calO(1/L^2)$ terms 
as discussed in section \ref{sec:LTBA}.
It is thus understood that the equations are valid up to relative
corrections, if any,  which are exponentially small in $A$ or of $ \calO(A^{-2})$.

\subsection{Cross-ratios}
\label{sec:CR}

\begin{figure}[t]
\vspace{5ex}
 \begin{center}
   \begin{minipage}{0.4\hsize}
   \begin{center}
  \includegraphics[width=70mm]{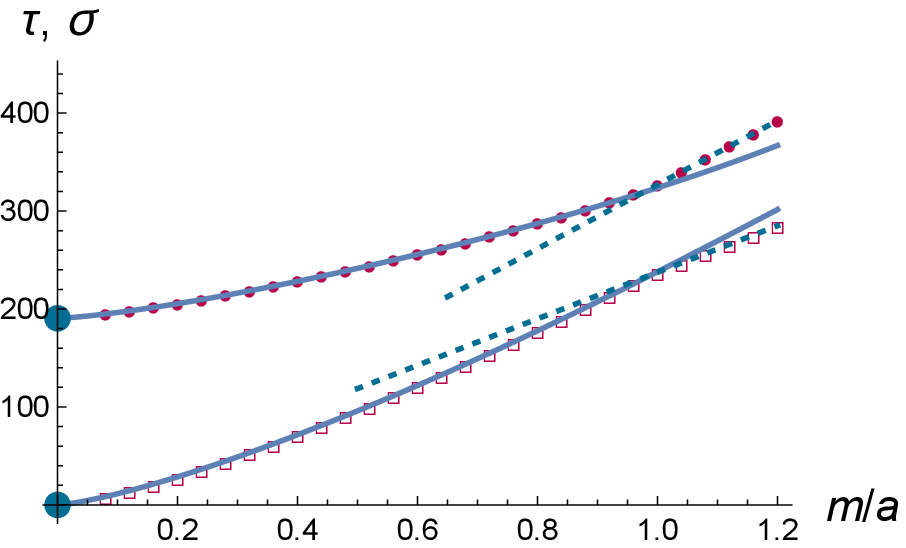}  \\ { \small (a)}
 \end{center}
 \end{minipage}
\hspace*{10ex}
\begin{minipage}{0.4\hsize}
 \begin{center}
  \includegraphics[width=70mm]{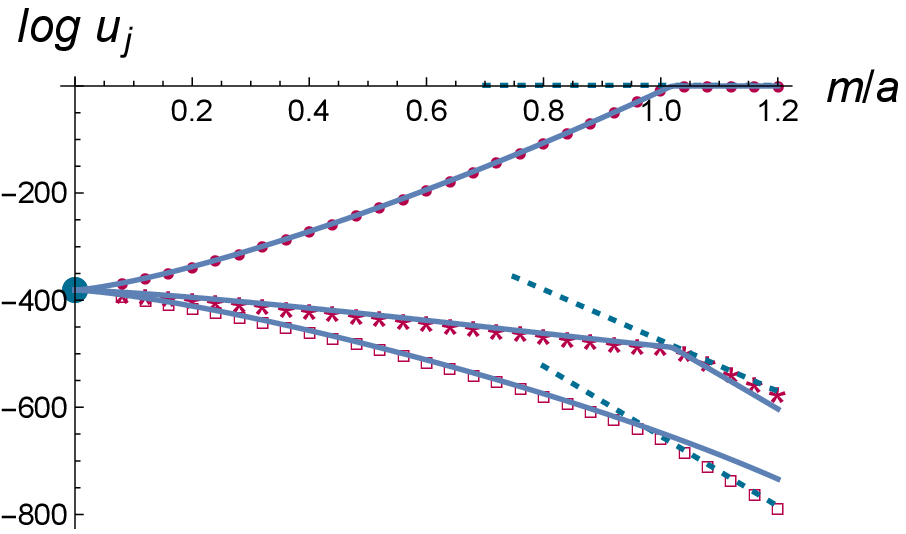}  \\ { \small (b)}
 \end{center}
 \end{minipage}
 \hfill
\caption{
 (a) Plot of  $\tau$ and $\sigma$.
The numerical data of $\tau$ and $\sigma$ are plotted by 
$\cdot$ and {\scriptsize $ \Box$}, respectively.
The solid lines represent the analytic expansions 
from those of $\varep, \tvarep$.
The blobs ($\bullet$) represent their UV values,
$\tau_{\rm UV}, \sigma_{\rm UV}$.
 The dotted lines represent their asymptotic forms, $\tau_{\rm IR}, \sigma_{\rm IR}$.
 The constant  $\phi = - A$ is omitted.
 (b) Plot of the cross-ratios  $u_j$.
 The numerical data of $u_1, u_2, u_3$ are plotted by $\cdot$,  {\scriptsize $ \Box$} 
 and $\ast$, respectively.
The solid lines represent the analytic expressions from those of $\varep, \tvarep$.
The blob ($\bullet$) represents the UV value.
The dotted lines represent the asymptotic forms in the IR region.
 }
 \label{fig:TauSigmaU}
\end{center}
\end{figure}

Let us first discuss how the cross-ratios change  
according to the changes in the parameters of the TBA  equations.
From the expansions (\ref{epexp2}) which are valid for any $\im \theta$,
one can obtain $u_j$ by substituting these into (\ref{ujUj}).
Alternatively,  one can consider the variables  in  (\ref{Ytsp}) parametrizing the cross-ratios.
From (\ref{Yep}) and  (\ref{Ytsp}), they are given by
\be
\label{ptsep}
\phi = -A \comma \quad  \tau = \frac{1}{2}\tvarep\Bigl( -\frac{i \pi }{4} - i\varphi \Bigr) 
  \comma \quad 
\sigma = A + \varep(-i\varphi) - \frac{1}{2} \tvarep\Bigl( -\frac{i \pi }{4} - i\varphi \Bigr)  
  \period
\ee
For small $m/a$, the pseudo energies $\ep, \tep$ approach 
 their UV values in (\ref{smallM}), and hence
\be
\label{tsUV}
\tau \to \tau_{\rm UV}  := \frac{A}{3}   
\comma \quad \sigma \to \sigma_{\rm UV} := 0 \period
\ee
For $m/a \gtrsim1$, the pseudo energies approach their IR forms (\ref{largeM}), 
and hence 
\be
\label{tsIR}
 \tau \to \tau_{\rm IR} :=  {M \o \sqrt{2}} \cos\bigl(\frac{\pi}{4} + \varphi  \bigr) 
 \comma \quad  
\sigma  \to \sigma_{\rm IR} :=  {M \o \sqrt{2}} \cos\bigl(\frac{\pi}{4} - \varphi  \bigr)  \period 
\ee
Figure \ref{fig:TauSigmaU}  (a) is a plot of 
$\tau$  and $\sigma$ for the same parameters 
of the TBA equations as in subsection \ref{subsec:madep}. 
The points ($\cdot$) and the boxes ({\scriptsize $ \Box$}) represent the numerical 
data of $\tau$ and $\sigma$,
respectively. 
The solid lines represent the analytic expansions 
from those of  $\varep, \tvarep$ in (\ref{epexp2}).
The UV values $\tau_{\rm UV} \sim 190$ and $\sigma_{\rm UV}=0$ are
denoted by the blobs ($\bullet$). 
 The asymptotic forms in the IR, 
$\tau_{\rm IR} \sim  327  \cdot (m/a)$,
$\sigma_{\rm IR} \sim  238 \cdot (m/a)$, are plotted by the dotted lines.
The other variable is a constant, $\phi = - A =  - (4/7) \cdot 10^{3} \sim - 571$.
 As $\varphi \to -\pi/4$, the IR form $\sigma_{\rm IR}$ is flattened, which corresponds
 to the collinear  limit.

%%%%%%%%
%  figures for cross-ratios
%%%%%%%%%
%
%%%%%

In the following, we consider the case where $-\pi/4 < \varphi < 0$, unless
otherwise stated,  as in the above.
Other cases are analyzed similarly.  In this case, we find that 
\be
  e^A \gg e^\tau  \gg e^\sigma \gg 1 \comma \nn
\ee
generically for $ 0 \leq m/a$.
The exceptions are $ e^\sigma \sim \calO(1)$ for small $m/a$ or 
$\varphi \to -\pi/4$, and
$e^{\tau-\sigma} \sim \calO(1)$ for $\varphi \sim 0$ and $m/a \gtrsim 1$.

The behavior of $(\tau, \sigma, \phi)$ is translated into that of $u_j$ by
using the relations (\ref{utsp}).
First, it follows from (\ref{tsUV}) that
\be
  u_1, u_2, u_3  \to u_{\rm UV} := e^{-\frac{2}{3}A} 
   \ll 1 \comma \nn
\ee
as $m/a \to 0$. To estimate $u_j$ for other $m/a$, we note that 
$ \sigma+\tau$ increases  as $m/a $ and approaches $M \cos \varphi$.
Thus, 
\be
\label{ujUV}
  u_1 \sim e^{\sigma+\tau-A} \comma \quad 
  u_2 \sim e^{-2\tau} \comma \quad 
  u_3 \sim e^{-\sigma+\tau-A} \comma
\ee
for $ m/a < 1$, where the expansions  
(\ref{epexp}) and (\ref{tepexp}), or (\ref{epexp2}) are valid. 
Their product is still a constant $u_1 u_2 u_3 \sim e^{-2A}$ approximately.
As $M$ reaches $A/\cos \varphi$, the leading term in $u_3$ changes and hence
\be
 \label{ujIR}
  u_1 \sim 1 -e^{A-\tau-\sigma} \comma \quad 
  u_2 \sim e^{-2\tau} \comma \quad 
  u_3 \sim e^{-2\sigma} \comma
\ee
for $m/a > 1/\cos\varphi$, where the asymptotic forms (\ref{tsIR}) are valid.
The cross-ratios $u_j$ are kept small as $m/a$ is varied.

Figure \ref{fig:TauSigmaU}  (b) is a plot of $u_j$. The points ($\cdot$),
boxes ({\scriptsize $ \Box$}) and asterisks ($\ast$)
represent  the numerical data of $u_1, u_2 $ and $u_3$, respectively,
for the same parameters as above.
They are obtained by substituting the numerical solution of (\ref{TBA6})
into the full expressions of the cross-ratios (\ref{utsp}) with (\ref{ptsep}).
The solid lines represent the corresponding analytic expansions 
from those of $\varep, \tvarep$ in (\ref{epexp2}).
The blob ($\bullet$) represents the value in the UV limit 
$\hat{u}_{\rm UV} \sim {u}_{\rm UV} = e^{-2A/3}$.
The dotted lines represent the asymptotic forms obtained from $\tau_{\rm IR}$ 
and $\sigma_{\rm IR}$.
Though the expansions are valid for $ m/a < 1$, the solid lines
approximate the numerical data even in the IR region.
They are further approximated by (\ref{ujUV}) for $m/a < 1/\cos \varphi$
and by  (\ref{ujIR}) for $m/a >1/ \cos \varphi$.

The trajectory of $u_j$ is also summarized in  Figure \ref{fig:Cross-ratios}
in section \ref{sec:RevAmplitudes}:
The UV point $(u_1, u_2, u_3) = (\hat{u}_{\rm UV}, \hat{u}_{\rm UV}, \hat{u}_{\rm UV})$ 
 for $m/a=0$ is  denoted by the lower blob in the figure. This corresponds to the
 regular polygonal limit.
As $m/a$ increases,
$u_j$ moves along the yellow surface $ u_1 u_2 u_3 \sim e^{-2A} $, 
until $m/a$ reaches $1/\cos \varphi$. The expansions 
(\ref{epexp}) and (\ref{tepexp}), or (\ref{epexp2}) are valid there for $m/a < 1$.
For larger $m/a$, the cross-ratios arrive at the IR point $(u_1, u_2, u_3) = (1,0,0)$
denoted by the upper blob.
This corresponds to a soft limit. 
The transition from (\ref{ujUV}) to (\ref{ujIR}) happens in a small region 
$\delta(m/a) \sim \calO(1/A)$.
This rapid change explains the apparent bends 
in Figure \ref{fig:TauSigmaU}  (b):
though some derivatives of the plots might look discontinuous around $m/a \sim 1$,
that is not the case.

In terms of $(\tau, \sigma, \phi)$, these two regimes are
smoothly connected. This is also confirmed by the fact that
 the yellow surface in Figure \ref{fig:Cross-ratios} asymptotes for large $A$  to
$u_1 u_2 u_3 =0$  which includes the IR end point.
Sending $\varphi \to -\pi/4$, one can adjust $u_3 \sim 1/(1+e^{2\sigma})$ for large enough
$m/a$, so that the end point becomes 
$(1-u_3, 0, u_3)$ with non-vanishing $u_3$. This corresponds to a collinear limit.
Thus, by changing $M$ and $\varphi$ (including generic $\varphi$),
the trajectories of $u_j$ for large $A$ form a  surface (``cap") which is close 
to  $u_1 u_2 u_3 = e^{-2A}$ but ends on the triangle of the collinear/soft limit.

\subsection{Expansion of 6-point remainder function}

We now focus on the region where $m/a < 1$ and the expansions of   
 (\ref{epexp}) and (\ref{tepexp}), or (\ref{epexp2}) are valid.
 The other region with $ m/a \gtrsim 1$ can be discussed  by using 
the asymptotic IR forms. In this region, $u_j$ are small, and the dilogarithm function
in  $\Delta A_{\rm BDS}$ in  (\ref{AperiodBDS6})
is approximated by ${\rm Li}_2(1-u_j^{-1}) \sim -\frac{1}{2}\log^2 u _j-\pi^2/6$.
Substituting the expansions (\ref{epexp2})
and omitting the terms relatively of $\calO(A^{-2})$
from $\pi^2/6$ above,
we  find  the $(m/a)$-expansion of $\Delta A_{\rm BDS}$: 
\be
 \label{expDelABDS}
  \Delta A_{\rm BDS} = A^2 \sum_{n=0}^\infty \beta_n y^{n/2} \comma
\ee
with
\begin{align}
\label{ABDSalpha}
  \beta_0 &= 
     \frac{1}{6} \comma &
  \beta_1 &= 0 \comma \nn  \\
  \beta_2 &= \frac{3\pi}{\Gamma^2(-\frac{1}{6}) \Gamma^2(\frac{2}{3})}  \comma  &
  \beta_3 &= -\frac{\cos(4\varphi)}{16\sqrt{3} \pi}  \comma &   \\     
 \beta_4 &= \frac{\pi}{48\Gamma^2(\frac{1}{6}) \Gamma^2(\frac{1}{3})} 
                       +  \frac{ 2^{\frac{1}{3}} \pi^{\frac{3}{2}} \Gamma(\frac{1}{6})}{
                       \Gamma^4(-\frac{1}{6}) \Gamma^2(\frac{2}{3})}   \comma   &
 \beta_5 &= -\frac{13\pi \cos(4\varphi)}{16\Gamma^2(-\frac{1}{3}) \Gamma^2(-\frac{1}{6})}  
 \comma &  \nn \\
 \beta_6 &= -\frac{17 \cos(8\varphi)}{2^9 3^3 \sqrt{3} \pi}  
                       +  \frac{19 \Gamma^2(\frac{1}{3}) \Gamma^6(\frac{7}{6})}{
                       128 \times 2^{\frac{1}{3}} \pi^4 \Gamma^2(\frac{5}{6}) }   \comma     & \nn                
\end{align}
for small $n$.
$\Delta A_{\rm BDS}$ is symmetric under $\varphi \to \varphi + 2\cdot ({\pi}/{4})$, which
permutes the three cross-ratios $u_j$. This symmetry strongly constrains the expansion, and
only the terms with $\cos\frac{4n}{3}\varphi$ $(n \in 3\bbZ)$
should survive \cite{Hatsuda:2010vr}.
From the expansions (\ref{epexp2}), such terms are not possible at  $\calO(y^{1/2})$.
The above expansion 
indeed satisfies this constraint with vanishing $\beta_1$.

 Combining the other terms, we obtain the expansion of the 6-point remainder function,
\eqb
  R_6  \Eqn{=}  \Delta A_{\rm BDS} - A_{\rm periods} - A_{\rm free}   \nn \\
     \Eqn{=} A^2 \sum_{n=0}^\infty \beta_n y^{n/2} 
    -   \frac{A^2}{\pi}  \sum_{n=0}^\infty k_n y^n \comma \nn 
\eqe
where $ \Delta A_{\rm BDS}$  is given in (\ref{expDelABDS}), $ A_{\rm periods} = M^2/4$
as in (\ref{AperiodBDS6}), and $A_{\rm free} = -L^2 f $ with $f$   given in
(\ref{fexp}). The coefficients $\beta_n$ and $k_n$ for small $n$ are given in   
(\ref{ABDSalpha}) and (\ref{fk}), respectively, whereas $y \sim (M/A)^{8\over 3}$ 
is defined in (\ref{fixingB}).
The bulk term $-M^2/4$ in $A_{\rm free}$ canceled  
$ A_{\rm periods}$.
The expansion is valid for $m/a <1$ and large $A=aL, M=mL$ up to the relative corrections
which are exponentially small  in $A$ or of $\calO(A^{-2})$. 

\begin{figure}[t]
\vspace{5ex}
 \begin{center}
   \begin{minipage}{0.4\hsize}
   \begin{center}
  \includegraphics[width=75mm]{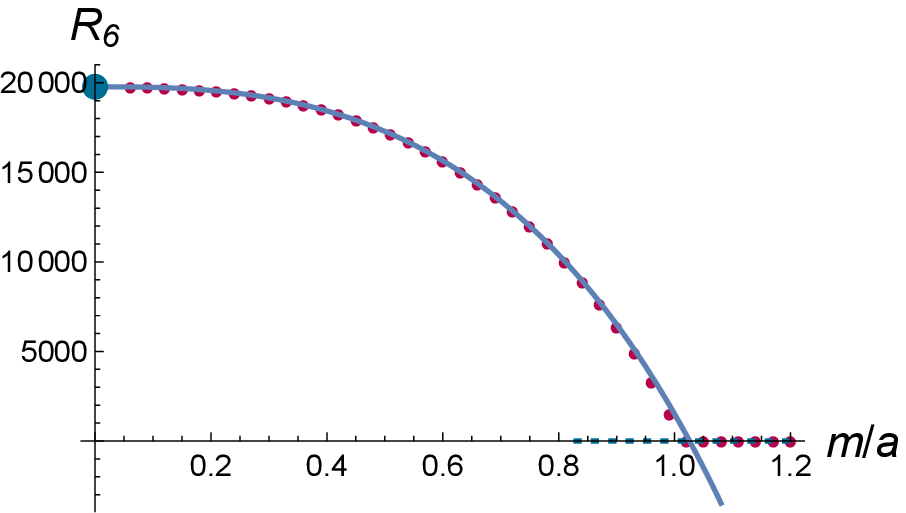}  \\ { \small (a)}
 \end{center}
 \end{minipage}
\hspace*{10ex}
\begin{minipage}{0.4\hsize}
 \begin{center}
  \includegraphics[width=70mm]{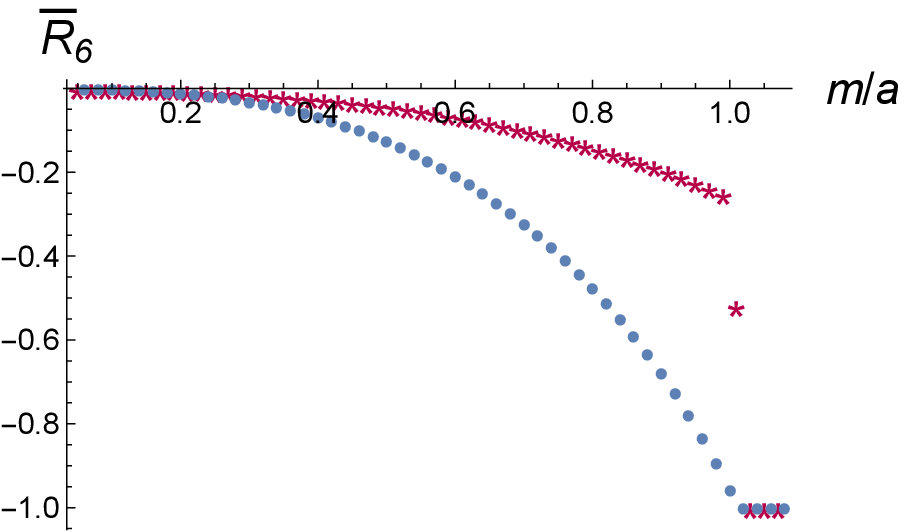}  \\ { \small (b)}
 \end{center}
 \end{minipage}
 \hfill
\caption{
 (a) Plot of 6-point remainder function at strong coupling.
 The points represent the numerical data. The solid line represents
 the analytic expansion. The blob ($\bullet$) represents  $R_{\rm 6, UV}$.
The dotted line represents  $R_{\rm 6, IR}$.
 (b) Plot of 6-point rescaled remainder functions. The points ($\cdot$)
 and the asterisks ($\ast$) represent the numerical data at strong coupling, 
 $\bar{R}_6^{\rm strong}$,
 and at 2 loops, 
 $\bar{R}_6^{\mbox{\scriptsize \rm 2-loop}} $, 
 respectively.
 }
 \label{fig:R6}
\end{center}
\end{figure}

Figure \ref{fig:R6} (a) is a plot of the 6-point remainder
function for the parameters as in 
subsection \ref{subsec:madep}.
The points ($\cdot$) represent the numerical data from the original TBA equations
(\ref{TBA6}). The solid line represents the expansion of $R_6$ given above
which includes terms up to $(m/a)^{8}$.
The blob ($\bullet$) is the value in the UV limit $R_{\rm UV}$ in (\ref{RUV}).
The dotted line represents the value  in the IR limit $R_{\rm IR}$ in (\ref{RIR}).

The expansion well approximates the numerical data  over the region from $m/a=0$
to $m/a = 1$, and interpolates the UV and the IR regime.
The relative correction of the expansion to the numerical value
 is $2.6 \times 10^{-5}$
at $m/a = 0.05$, and $2.5 \times 10^{-2}$ at $m/a = 0.9$. The former is attributed 
to the $\calO(L^{-2})$ correction in our expansions, whereas the latter is 
to the truncation of the expansions. 
By including higher order terms, one obtains a better approximation to the numerical data
 up to $m/a \sim 1$.
If we use the expression of $\Delta A_{\rm BDS}$ 
by substituting
the expansions (\ref{epexp2}) into (\ref{AperiodBDS6}) (without expanding in $y^{n/2}$),
higher order terms are partially incorporated and its IR behavior for  $ m/a \gtrsim 1$
is  improved.

%%%%%%%%
%  figures for remainder function
%%%%%%%%%
%

%%%%%
%
\subsection{Comparison with 2-loop results}

In \cite{Brandhuber:2009da,Hatsuda:2011ke,Hatsuda:2011jn},  
the remainder functions corresponding to the minimal surfaces in AdS${}_3$
were compared at strong coupling and at 2 loops \cite{DelDuca:2009au,Goncharov:2010jf} 
by changing the mass parameters
of the TBA system with other parameters fixed. It was found that they are close to each other 
(but different) after appropriately rescaled/normalized.
Such an observation was also made for the AdS${}_4$ case \cite{Hatsuda:2012pb}.
In the present 6-point case with the general  (AdS${}_5$) kinematics,
the comparison was extended and the similarity was confirmed up to 4 loops 
\cite{Hatsuda:2014vra}
based on the perturbative results \cite{Dixon:2013eka,Dixon:2014voa}.
The comparisons 
for other kinematics are found  in \cite{Dixon:2013eka,Dixon:2014voa}.
In this subsection, we compare the remainder functions at strong coupling 
and at 2 loops for the cross-ratios discussed in the previous subsections.

For this purpose, we introduce the remainder function at strong coupling
which is rescaled by its UV and IR values,
\be
  \bar{R}_6^{\rm strong} := \frac{R_6 - R_{\rm 6, UV}}{ R_{\rm 6, UV} - R_{\rm 6, IR} } \comma
  \nn
\ee
with $R_{\rm 6, UV}, R_{\rm 6, IR}$ given in (\ref{RUV}) and (\ref{RIR}), respectively.
For the remainder function in perturbation $R_6^{\rm pert} = \sum \lambda^\ell R_6^{(\ell)}$
where $\lambda$ is the 't Hooft coupling, the rescaled remainder functions 
$\bar{R}_6^{\ell\mbox{\scriptsize \rm -loop}} $
are defined similarly. 
In particular at 2 loops, 
\be
   \bar{R}_6^{\mbox{\scriptsize \rm 2-loop}} 
  := \frac{R_6^{(2)} - R^{(2)}_{\rm 6, UV}}{ R^{(2)}_{\rm 6, UV} - R^{(2)}_{\rm 6, IR} } \comma
  \nn
\ee
where
$R^{(2)}_{\rm 6,UV} = R^{(2)}_{6}(\hat{u}_{\rm UV},\hat{u}_{\rm UV},\hat{u}_{\rm UV})$, 
$R^{(2)}_{\rm 6,IR} = 0$. The analytic expression of $R^{(2)}_{\rm 6} $ is given 
\cite{DelDuca:2009au,Goncharov:2010jf}
by
\eqb
\label{R62loop}
 R_6^{(2)}(u_1,u_2,u_3) \Eqn{=} \sum_{i=1}^3 \left( L_4(x_i^+, x_i^-) -\frac{1}{2} {\rm Li}_4(1-u_i^{-1}) \right)
 -\frac{1}{8} \left( \sum_{i=1}^3 {\rm Li}_2(1-u_i^{-1})\right)^2 \nn \\
  && \ + \ \frac{1}{24} J^4 + \frac{\pi^2}{12} J^2+ \frac{\pi^4}{72}  \comma
\eqe
where
\be
  x_i^\pm = u_i x^\pm \comma \quad
  x^\pm = \frac{u_1 + u_2 + u_3 -1 \pm \sqrt{\Delta}}{2 u_1 u_2 u_3} \comma \quad
  \Delta = (u_1+u_2+u_3-1)^2 - 4u_1 u_2 u_3 \comma \nn
\ee
and
\eqb
  L_4(x^+,x^-) \Eqn{=} \frac{1}{8!!} \log^4(x^+ x^-) 
   + \sum_{m=0}^3 \frac{(-1)^m}{(2m)!!} \log^m(x^+x^-) \times \bigl(\ell_{4-m}(x^+) 
   + \ell_{4-m}(x^-) \bigr)
   \comma \nn \\
  \ell_n(x) \Eqn{=} \frac{1}{2} \left( {\rm Li}_n(x) -(-1)^n {\rm Li}_n(x^{-1}) \right) \comma \nn \\
   J \Eqn{=} \sum_{i=1}^3 \left( \ell_1(x_i^+) - \ell_1(x_i^-) \right) \period \nn
\eqe
By definition, $\bar{R}_6^{\rm strong}, \bar{R}_6^{\mbox{\scriptsize \rm 2-loop}} \to 0$ 
in the UV limit, and 
$\bar{R}_6^{\rm strong}, \bar{R}_6^{\mbox{\scriptsize \rm 2-loop}}  \to -1$ 
in the IR limit.

Figure \ref{fig:R6} (b) is a plot of the rescaled remainder functions at strong coupling 
and at 2 loops for the cross-ratios given in section \ref{sec:CR}.
The points $(\cdot)$ represent the numerical data of 
$\bar{R}_6^{\rm strong}$ which are obtained 
from the numerical solution of (\ref{TBA6}). 
The asterisks ($\ast$) represent 
$\bar{R}_6^{\mbox{\scriptsize \rm 2-loop}} $
which are obtained by substituting into (\ref{R62loop}) the same numerical values 
of the cross-ratios.

To avoid loss of significant digits in the numerical evaluation of $R^{(2)}_6$, 
we have made an approximation,
\be
  x^+ \sim \frac{1}{u_1+u_2+u_3-1}   \comma \qquad
  x^- \sim \frac{u_1+u_2+u_3-1}{u_1 u_2 u_3 }  \comma \nn
\ee
which is valid for $ (1-u_1 -u_2 - u_3)^2  \gg  u_1 u_2 u_3$, and hence both for 
(\ref{ujUV}) and (\ref{ujIR}) up to exponentially small terms in $A$.
By using the analytic expansion of $R_6$ at strong coupling, one can draw 
a plot of $\bar{R}_6^{\rm strong}$ which well approximates the data points as in 
Figure \ref{fig:R6} (a). 
Since the cross-ratios $u_j$ are well approximated by the analytic expansions,
substituting them into $R_6^{(2)}$  in (\ref{R62loop}) also gives a plot 
which approximates the data points as well.

The behaviors of $\bar{R}_6^{\rm strong}$  and 
$\bar{R}_6^{\mbox{\scriptsize \rm 2-loop}} $
are rather different
for large $m/a$, where $\bar{R}_6^{\rm strong}$ decreases faster than 
$\bar{R}_6^{\mbox{\scriptsize \rm 2-loop}} $.
This is in contrast to the results mentioned above 
by varying the mass parameters with others fixed.

%%%
\section{Conclusions}

Through the gauge-string duality, the MHV amplitudes of ${\cal N}=4$ SYM 
at strong coupling are obtained by solving  auxiliary integral equations of the TBA type. 
In this paper, we considered the limit where  chemical potentials in addition to  masses
are large, and hence the TBA equations 
are linearized. Large chemical potentials, together with large masses,
 thus  provide another useful limit. In particular, 
the linearized TBA equations for the 6-point amplitudes 
are solved analytically according to \cite{AlZamolodchikov1995}
as expansions to any order in terms of the ratio of the mass and the chemical potential $M/A$. 
The relative corrections in the linearization are exponentially small in $A$ or of $\calO(1/A^2)$.
The inverse power corrections are analyzed by 
extending the analysis in \cite{AlZamolodchikov1995}, and we obtained the explicit forms
of the leading corrections.
We checked that our results agree with numerical solutions. 

Assured that the linearization
gives a controlled approximation of the original TBA equations for large $A$ and $M$, 
we derived analytic expansions of the 6-point MHV amplitudes
from the solution of the linearized TBA equations.
The expansion is  again valid to any order
up to corrections exponentially small in $A$ or of $\calO(A^{-2})$.
As $M$  is varied with $A \gg 1$ and the phase $\varphi$ fixed, the three cross-ratios $u_j$ 
of the momenta of scattering particles are kept small and 
change from the equal value 
$u_j  \sim e^{-2A/3} $ in the UV regime to those in the soft/collinear limits in the IR regime. 
The amplitudes are well described by the expansion over the corresponding kinematic 
region.

We also compared the 6-point rescaled remainder functions at strong coupling and 
at 2 loops along the trajectory of $u_j$ mentioned above.
We observed that they are rather different, 
in contrast to the  cases where similarities are observed between the strong-coupling 
results and the perturbative results  \cite{Brandhuber:2009da,Hatsuda:2011ke,Hatsuda:2011jn,Hatsuda:2012pb,Hatsuda:2014vra,Dixon:2013eka,Dixon:2014voa}.
This implies that the kinematic region of the small cross ratios  provides a 
 useful probe to study structural differences of the strong-coupling and the perturbative
results. For example, since the perturbative results share the property that they are 
controlled by the transcendentality and the associated symbol 
\cite{Goncharov:2010jf,Dixon:2013eka,Dixon:2014voa,Caron-Huot:2016owq}, 
the difference from the strong-coupling case 
may persist for higher loops.
The actual comparison with higher-loop results in 
\cite{Dixon:2013eka,Dixon:2014voa,Caron-Huot:2016owq}
is thus an interesting future problem.

The extension of our analysis to the general $n$-point amplitudes would be 
an important future direction. For $n >6$, the TBA system has more than one mass scale,
and the generalization of the work \cite{AlZamolodchikov1995} is not straightforward.
This is closely related to the problem of finding the exact mass-coupling relation, 
i.e. the relation between the physical mass and the coupling in the Lagrangian, for multi-scale
integrable models. 
 As mentioned in section \ref{sec:RevAmplitudes},
this problem has been solved  \cite{Bajnok:2015eng,Bajnok:2016ocb} 
for  a simple multi-scale integrable model, i.e. the $su(3)_2/u(1)^2$ HSG model, 
which is relevant for the 10-point amplitudes for two-dimensional kinematics.
The issue of the multi-scales has been overcome there by comparing 
the picture of the conformal perturbation on the UV side and that of the form-factor 
bootstrap on the IR side.

By adjusting the phase $\varphi$, the IR end point of the trajectory of 
$u_j$
can be set to the point of the collinear limit, around which the OPE expansion at finite coupling
is derived \cite{Alday:2010ku,Basso:2013vsa}. It would be of interest to consider
if our expansion at strong coupling  provides useful data for the OPE expansion.

%%%%%%%%%%%%%%%%%%%%%%%%%%%%%%%%%%
\par\bigskip
\par\bigskip
\begin{center}
{\large\bf Acknowledgments}
\end{center}
Y.S. would like to thank
Zoltan Bajnok for useful discussions and for sharing his notes on the linearization 
of TBA equations, and  Davide Fioravanti for useful discussions. 
This work is supported in part by JSPS Grant-in-Aid for Scientific Research 
24540248, 15K05043, 15K05208, 16F16735, 17K05406, 18H01141, 18K03452 and 18K03643, 
and Japan-Hungary Research Cooperative Program 
from Japan Society for the Promotion of Science (JSPS).
%%%%%%%%%%%%%%%%%%%%
%%%%%%%
\par\bigskip
\par\bigskip

\appendix
\renewcommand{\theequation}{\Alph{section}.\arabic{equation}}

\section{Evaluation of $\tep$ from direct integrals }
\label{app:tep}

In this appendix, we evaluate the expansion of $\tep$ in (\ref{approx1epstilde})
by  plugging that of $\ep$ (\ref{approx1eps}) into (\ref{eq:tba2dd}). 
For this purpose, we note the formula, 
\be
  \int dx \, \frac{e^{cx}}{\cosh(ax+b)} = \frac{2}{a+c} e^{(a+c)x+b} 
  {}_2F_1\Bigl(1,\frac{a+c}{2a}; 1+\frac{a+c}{2a};-e^{2(ax+b)} \Bigr) \period
    \label{intcosh}
\ee
This is checked by the expansion of the  hypergeometric function,
\be
 {}_2F_1(1,\beta;\beta+1;z) = \sum_{n=0}^\infty \frac{\beta z^n}{\beta+n} =: \Phi_\beta(z)
 \comma
  \nn
\ee
for $|z|$  $< 1$, 
and its analytic continuation in $z$, from which we find
\be
 \Phi_\beta(z) + \frac{z}{\beta}\Phi'_\beta(z) = \frac{1}{1-z} \period 
  \label{dPhi}
\ee
Taking the derivative of the right hand side of (\ref{intcosh}) with the help of (\ref{dPhi}) and setting
$z=-e^{2(ax+b)},\beta={(a+c)/2a}$, we obtain the formula.
Furthermore using the inversion formula,
\begin{align}
\label{F21inv}
  {}_2F_1(a,b;c;z)  &=  
  \frac{ \Gamma(b-a) \Gamma(c)}{\Gamma(b) \Gamma(c-a)}  (-z)^{-a} {}_2F_1(a,a-c+1;a-b+1;z^{-1})
   + ( a \leftrightarrow b) 
  \comma 
\end{align} 
and 
\begin{align}
  {}_2F_1(a,b;b;z) & =  (1-z)^{-a}  \comma \qquad \Gamma(1-z) \Gamma(z) = \frac{\pi}{\sin \pi z}
  \comma \nn
\end{align}
one finds
\be
   \int_{-B}^B d\theta' \, \frac{e^{c\theta'}}{\cosh2(\theta'-\theta)} =  \frac{\pi}{2} \frac{e^{c\theta}}{\sin \pi \beta} 
  - \frac{e^{-2\theta - 4\beta B}}{2\beta} \Phi_\beta\bigl(-e^{-4(B+\theta)} \bigr)
  - \frac{e^{2\theta - 4(1-\beta) B}}{2(1-\beta)} \Phi_{1-\beta}\bigl(-e^{-4(B-\theta)} \bigr)
  \comma \nn
\ee
with $\beta = (2+c)/4$.  
Using this, we get
\begin{align}
&  I_n(B, \theta) := 
\int_{-B}^B \frac{d\theta'}{2\pi} K_2(\theta-\theta')  \cosh\Bigl( \frac{4n}{3}\theta'\Bigr)
          \nn  \\
  & \qquad   {=} \, (-1)^n \cosh\Bigl(\frac{4n}{3} \theta \Bigr)
  - \frac{\sqrt{2}}{2\pi} \sum_{\ell,\ell'=\pm 1}
  \frac{e^{-(2+s_n)B} }{2+s_n} 
 \Bigl[  e^{(2+\ell)\theta} \Phi_{\frac{2+s_n}{4}} \bigl(-e^{-4(B-\theta)} \bigr) 
 + (\theta \to - \theta) \Bigr]
  \comma \nn
\end{align}
with $s_n = \frac{4n}{3} \ell' + \ell$. From (\ref{approx1eps}) and 
(\ref{eq:tba2dd}), we thus find 
\be
  \tep(\theta) = \sqrt{2} m\cosh \theta + \frac{2a}{3} I_0(B,\theta)  
  + \sum_{n>0 \, n\neq0 \, ({\rm mod} \, 3)} c_n I_n(B,\theta) \period \nn
\ee 
Comparing this with (\ref{approx1epstilde}), we observe that
the summation over $ \Phi_{(2+s_n)/4}$ in $I_n$ cancels the driving term
$\sqrt{2} m \cosh \theta$, which leaves the first terms $(-1)^n \cosh\bigl({4n} \theta/3 \bigr) $
in accordance with  the periodicity required from the Y-system.
These fractional powers of $e^\theta$ appeared by summing up integral powers of $e^{\theta}$
in $\Phi_\beta$ and analytically continuing it by the inversion (\ref{F21inv}).

%%%
\section{Estimation of $ \epsilon'(B)$ }
\label{app:gamma1}
We consider the evaluation of $\epsilon'(B)$ (and $\xi_1$), which is necessary for 
$\kappa_{nm}$  in (\ref{cdnell}) and $q^{(1)}$ in (\ref{q01}) for example.
For the derivative of (\ref{linearALz}) with respect to $\theta$, we obtain
\begin{equation}
D_{\epsilon'}(\theta) = \int_{-\infty}^{\infty}  \tilde{K}(\theta-\theta') \epsilon'_B(\theta') d\theta',
\nn
\end{equation}
where 
\begin{equation}
D_{\epsilon'}(\theta)  =y(\theta)-y(-\theta),
\qquad
y(\theta)=  \begin{cases}
 -\frac{m}{2} {\rm e}^{\theta} &    \theta<B \\
 x(\theta)&  \theta>B .
\end{cases}
\nn
\end{equation}
Here  the terms like $\Delta_1'(\theta)$ are neglected as they are higher order terms.

Set 
\[
\tau(w)= \Km (w) \widehat{y}(w) {\rm e}^{-iwB}.
\]
By repeating  the same argument as in section \ref{sec:LTBA}, 
we obtain an integral equation for $\tau(w)$,
\begin{align*}
\tau(w) &=D_{\tau}(w)-\int_{-\infty}^{\infty} \frac{ {\rm e}^{2iw'B}}{w+w'+i\varepsilon} \alpha(w') \tau(w')  \frac{dw'}{2\pi i},
 \\
&D_{\tau}(w)= i \frac{m e^{B} \Kp(-i)}{2(w-i)}.  
\end{align*}
The solution is formally given by
\begin{equation}\label{eq:tau_expansion1}
\tau(w)=  i \frac{m e^{B} \Km(i)}{2(w-i)}
- \sum_{n\ge 1, n\ne 0 ({\rm mod} 3)} \frac{ {\rm e}^{-\frac{8}{3} n B}}{w+\frac{4}{3}n i} \alpha_n  \tau(\frac{4}{3}n i), 
\end{equation} 
where  $\alpha_n = {\rm res}_{w=\frac{4}{3}n i} \alpha(w)$.

Thus we obtain a set of algebraic equations for $\hat{\tau}_\ell$ ($\ell \in \mathbb{N}$),
\begin{align*}
\widehat{\tau}_{\ell} &= \frac{1}{\frac{4{\ell}}{3}-1} -
 \sum_{n\ge 1, n\ne 0 ({\rm mod} 3)}  \frac{ q^n }{{\ell}+n}
 \frac{(-1)^n}{(n-1)!n!} 
 \frac{\Gamma(1+\frac{n}{3}) \Gamma(\frac{1}{2}+\frac{2n}{3}) }
   {\Gamma(1-\frac{n}{3}) \Gamma(\frac{1}{2}-\frac{2n}{3}) }\,\widehat{\tau}_n, \\
 & \widehat{\tau}_{\ell} =\tau(\frac{4}{3}i{\ell}) \frac{2 {\rm e}^{-B}}{m\Km(i)}.
\end{align*}
From this, one obtains $\widehat{\tau}_m$ as a power series in $q$, 
$ \widehat{\tau}_{\ell} = \frac{1}{\frac{4 \ell }{3}-1} + \calO(q)$.
By substituting them into (\ref {eq:tau_expansion1}) we obtain $\tau(w)$.
Then $\widehat{\epsilon'}_B(w)$ is evaluated by
\[
\widehat{\epsilon'}_B(w) =\Kp(w) \tau(w) {\rm e}^{iwB} -  \Km(w) \tau(-w) {\rm e}^{-iwB}.
\]
The inverse Fourier transformation yields $\epsilon'_B(\theta)$,
\be
\label{DepB}
\epsilon'_B(\theta ) =\int_{-\infty}^{\infty}  \frac{dw}{2\pi} {\rm e}^{iw(B-\theta)}  \Kp(w) \tau(w) 
-\int_{-\infty}^{\infty}  \frac{dw}{2\pi} {\rm e}^{iw(B+\theta)}  \Kp(w) \tau(w).
\ee 
In particular, at $\theta=B$, 
we deform in the first term the finite part of the integration contour 
to a semi-circle in the lower half plane, and take the large radius limit of the semi-circle. 
Then, we obtain
\be
\label{DepB1}
  \int_{-\infty}^{\infty}  \frac{dw}{2\pi}  \Kp(w) \tau(w)  
  = -  \frac{m}{4} \Km(i) {\rm e}^B 
 \biggl[ 1 + i \sum_{ \substack{n\ge 1\\ n\ne 0 ({\rm mod} 3)} } 
 {\rm e}^{-\frac{8}{3} n B}  \alpha_n  \widehat{ \tau}_n 
  \Bigl(  1-2  \Kp\bigl( -\frac{4}{3}n i \bigr) \Bigr)
\biggr]. 
\ee
The first and the second term in the bracket 
come from the semi-circle, whereas the third from the poles  
which are picked up in deforming the contour.
We have used $\Kp(w) \to 1 $ for large $|w|$. 
In the second term in (\ref{DepB}), we close the integration contour  
in the upper half plane, to obtain the contributions which cancel the third term 
 in (\ref{DepB1}). Combining these,  we find
\be
 \label{DepB2}
 \epsilon'_B(B) = -  \frac{m}{4} \Km(i) {\rm e}^B
  \biggl[ 1 + i \sum_{ \substack{n\ge 1\\ n\ne 0 ({\rm mod} 3)} } 
 {\rm e}^{-\frac{8}{3} n B}  \alpha_n  \widehat{ \tau}_n  \biggr]. 
\ee

We note  $\epsilon'_B(B) =   \frac{1}{2}\bigl( \epsilon'_B(B-0^+)  + \epsilon'_B(B+0^+)\bigr)
=   \frac{1}{2} \epsilon'_B(B-0^+)$, as
$\epsilon'_B(\theta) $ is discontinuous at $|\theta| = B$ and vanishes for $|\theta|>B$.
Thus, $ |\epsilon'(B)|  =  |\epsilon_B'(B-0^+)| = 2 |\epsilon_B'(B)| $ with 
$\epsilon_B'(B)$ in  (\ref{DepB2}).
Since $B=\calO(L^0)$,  it is clear that $\epsilon'(B)=\calO(L^0)$.  

%%

%\newpage 
\par\medskip
\par\bigskip
%%%%%%%
\small

\setlength{\parskip}{0ex}

\end{document}